%% file: main.tex
\documentclass[a4paper, preprint]{cas-dc}
\usepackage[numbers]{natbib}

\let\printorcid\relax 

\input{header} 

\title[mode=title]{Integrating AI in NDE: Techniques, Trends, and Further Directions}

\shorttitle{Integrating AI in NDE: Techniques, Trends, and Further Directions}
\shortauthors{Eduardo Pérez et~al.}

\author[1,2]{Eduardo Pérez}
\credit{Conceptualization of this study, Methodology, Software}

\author[1,2]{Cemil Emre Ardic}
\author[1,2]{Ozan Çakıroğlu}
\author[1,3]{Kevin Jacob}
\author[1]{Sayako Kodera}
\author[1]{Luca Pompa}
\author[1]{Mohamad Rachid}
\author[1,2]{Han Wang}
\author[1,4]{Yiming Zhou}
\author[1]{Cyril Zimmer}
\author[1]{Florian Römer}
\author[1,4]{Ahmad Osman}

\affiliation[1]{organization={Fraunhofer-Institut für Zerstörungsfreie Prüfverfahren IZFP},
    addressline={Campus E3 1},
    city={66123 Saarbrücken},
    country={Germany}}
\affiliation[2]{organization={Technische Universität Ilmenau},
    addressline={Ehrenbergstraße 29},
    city={98693 Ilmenau},
    country={Germany}}
\affiliation[3]{organization={Universität des Saarlandes},
    addressline={Campus},
    city={66123 Saarbrücken},
    country={Germany}}

\affiliation[4]{organization={Hochschule für Technik und Wissenschaft des Saarlandes},
    addressline={Goebenstrasse 40},
    city={66117 Saarbrücken},
    country={Germany}}
    
\date{}

\begin{document}
\let\WriteBookmarks\relax
\def\floatpagepagefraction{1}
\def\textpagefraction{.001}

\begin{keywords}
Artificial intelligence \sep Machine Learning \sep NDE 4.0 \sep ECT \sep 3MA \sep UT \sep TT \sep VT
\end{keywords}

\input{sections/abstract}

\maketitle

\input{sections/introduction}
\input{sections/3ma}

\input{sections/ultrasound}
\input{sections/thermography}

\input{sections/optical_camera}
\input{sections/overview_shrink}
\input{sections/summary_conclusions}

\input{sections/closing}
\vspace{-0.3cm}


\setglossarystyle{index} 
\raggedright             
\printglossaries

\bibliographystyle{unsrt_initialfirstnames}
\bibliography{references}
\end{document}

%% file: header.tex
\usepackage[OT2,T1]{fontenc} %
\DeclareSymbolFont{cyrletters}{OT2}{wncyr}{m}{n}
\DeclareMathSymbol{\Sha}{\mathalpha}{cyrletters}{"58}

\usepackage{amsmath,amsthm,amssymb,amsfonts}   
\usepackage{bm}                         
\usepackage{graphicx}                   
\usepackage{psfrag}                     
\usepackage{bbm}						
\usepackage[font=small]{caption}
\usepackage{subcaption}
\usepackage{todonotes}
\usepackage{url}
\usepackage{placeins}
\usepackage{textcomp}
\usepackage{siunitx} 
\usepackage[stretch=40,shrink=40]{microtype}
\usepackage{tikz}
\usepackage{ifthen}
\usetikzlibrary{patterns}
\usetikzlibrary{shapes, arrows, chains, fit, positioning, calc, decorations.pathreplacing}
\usetikzlibrary{petri}
\usepackage{pgfplots}
\pgfplotsset{compat = 1.14}
\usepgfplotslibrary{colormaps}
\usepgfplotslibrary{groupplots}
\usepackage{multicol} 
\usepackage[dvipsnames]{xcolor}
\usepackage{nicematrix} 
\usepackage{comment}
\usepackage{lipsum}
\usepackage{cite} 
\usepackage[nameinlink]{cleveref}
\usepackage[normalem]{ulem}
\usepackage{hyperref}
\usepackage[automake, acronym, nopostdot, nonumberlist, nogroupskip]{glossaries} 

\newcommand{\real}{\mathbb{R}}          
\newcommand{\herm}{^{\text{H}}}			
\newcommand{\red}[1]{\textcolor{red}{#1}} 
\newcommand{\cy}[1]{\textcolor{Fuchsia}{#1}} 

\DeclareMathOperator*{\argmin}{argmin}

\newdefinition{rmk}{Remark}
\newproof{pf}{Proof}
\newproof{pot}{Proof of Theorem \ref{thm2}}

\input{tikz_tables/tikzstyles.tex}

\newcommand{\rom}[1]{\uppercase\expandafter{\romannumeral #1\relax}}

\makeglossaries
\glsdisablehyper 
\newacronym{ndt}{NDE}{Nondestructive Evaluation} 
\newacronym{nde}{NDE}{Nondestructive Evaluation} 

\newacronym{adr}{ADR}{Automated Defect-Recognition}

\newacronym{ct}{CT}{Computed Tomography}
\newacronym{uct}{UCT}{Ultrasound Computed Tomography}
\newacronym{us}{US}{Ultrasound}
\newacronym{ut}{UT}{Ultrasonic Testing}
\newacronym{ect}{ECT}{Eddy Current Testing}
\newacronym{emat}{EMAT}{Electro Magnetic Acoustic Transducer}
\newacronym{3ma}{3MA}{Micromagnetic Multiparameter Microstructure and stress Analysis}
\newacronym{pwi}{PWI}{Plane Wave Imaging}

\newacronym{uq}{UQ}{Uncertainty Quantification}

\newacronym{ai}{AI}{Artificial Intelligence}
\newacronym{ml}{ML}{Machine Learning}
\newacronym{mlp}{MLP}{Multilayer Perceptron}
\newacronym{nn}{NN}{Neural Network}
\newacronym{ann}{ANN}{Artificial Neural Network}
\newacronym{dnn}{DNN}{Deep Neural Network}
\newacronym{cnn}{CNN}{Convolutional Neural Network}
\newacronym{rnn}{RNN}{Recurrent Neural Network}
\newacronym{resnet}{resNET}{Residual Neural Network}
\newacronym{gan}{GAN}{Generative Adversarial Neural Network}
\newacronym{dcgan}{DCGAN}{Deep Convolutional Generative Adversarial Network}
\newacronym{gmdh}{GMDH}{Group Method of Data Handling}
\newacronym{svm}{SVM}{Support Vector Machine}
\newacronym{svr}{SVR}{Support Vector Regression}
\newacronym{hlr}{HLR}{Huber Loss Regression}
\newacronym{rbf}{RBF}{Radial Basis Function}
\newacronym{knn}{$k$NN}{$k$-Nearest Neighbour}
\newacronym{lda}{LDA}{Linear Discriminant Analysis}
\newacronym{rf}{RF}{Random Forest}
\newacronym{xgboost}{XGBoost}{extreme Gradient Boosting}
\newacronym{pinn}{PINN}{Physics-Informed Neural Network}
\newacronym{mlr}{MLR}{Multiple Linear Regression}
\newacronym{kfcv}{$k$FCV}{$k$-Fold Cross Validation}
\newacronym{pca}{PCA}{Principle Component Analysis}
\newacronym{rcnn}{R-CNN}{Region Based Convolutional Neural Network}
\newacronym{iiot}{IIoT}{Industrial Internet of Things}

\newacronym{r2}{\textit{R}\textsuperscript{2}}{Coefficient of Determination}
\newacronym{cor}{\textit{R}}{Correlation Coefficient}
\newacronym{mse}{MSE}{Mean Squared Error}
\newacronym{rmse}{RMSE}{Root Mean Squared Error}
\newacronym{mae}{MAE}{Mean Absolute Error}
\newacronym{mle}{MLE}{Maximum Likelihood Estimator}
\newacronym{sd}{SD}{Standard Deviation}
\newacronym{pod}{PoD}{Probability of Detection}

\newacronym{pde}{PDE}{Partial Differential Equation} 

\newacronym{ista}{ISTA}{Iterative Shrinkage-Thresholding Algorithm}
\newacronym{fista}{FISTA}{Fast ISTA}
\newacronym{admm}{ADMM}{Alternating Direction Method of Multipliers}
\newacronym{sgd}{SGD}{Stochastic Gradient Descent}
\newacronym{snr}{SNR}{Signal to Noise Ratio}
\newacronym{psnr}{PSNR}{Peak Signal-to-Noise Ratio}
\newacronym{dps}{DPS}{Deep Probabilistic Subsampling}
\newacronym{jdps}{J-DPS}{Joint DPS}

\newacronym{rpv}{RPV}{Reactor Pressure Vessel}
\newacronym{lcf}{LCF}{Low-cycle fatigue}
\newacronym{dbtt}{DBTT}{Ductile-Brittle Transition Temperature}
\newacronym{mat}{MAT}{Magnetic Adaptive Testing}
\newacronym{dcrpd}{DCRPD}{Direct Current Reversal Potential Drop}
\newacronym{tepmm}{TEPMM}{Thermoelectric Power Measuring Method}
\newacronym{mirbe}{MIRBE}{Micromagnetic Inductive Response and Barkhausen Emission Method}

\newacronym{re}{\textit{R}\textsubscript{e}}{Yield Strength}
\newacronym{rm}{\textit{R}\textsubscript{m}}{Tensile Strength}

\newacronym{tfm}{TFM}{Total Focusing Method}
\newacronym{fem}{FEM}{Finite Element Method}
\newacronym{fe}{FE}{Finite Element}
\newacronym{fmc}{FMC}{Full Matrix Capture}
\newacronym{saft}{SAFT}{Synthetic Aperture Focusing Technique}
\newacronym{fwhm}{FWHM}{Full-Width at Half Maximum}
\newacronym{das}{DAS}{Delay and Sum}
\newacronym{pw}{PW}{Plane Wave}
\newacronym{cnr}{CNR}{Contrast-to-Noise Ratio}
\newacronym{drl}{DRL}{Deep Residual Learning}
\newacronym{lista}{LISTA}{Learned ISTA}
\newacronym{mv}{MV}{Minimum Variance}
\newacronym{able}{ABLE}{Adaptive Beamforming by Deep LEarning}
\newacronym{admire}{ADMIRE}{Aperture Domain Model Image REconstruction}
\newacronym{ceus}{CEUS}{Contrast-Enhanced Ultrasound}
\newacronym{corona}{CORONA}{Convolutional Robust Principal Component Analysis}
\newacronym{cs}{CS}{Compressed Sensing}
\newacronym{coba}{COBA}{Convolutional Beamforming Algorithms}
\newacronym{fri}{FRI}{Finite Rate of Innovation}
\newacronym{art}{ART}{Algebraic Reconstruction Technique}
\newacronym{sirt}{SIRT}{Simultaneous Iterative Reconstruction Technique}
\newacronym{fwi}{FWI}{Full Waveform Inversion}
\newacronym{dn}{DN}{Dense Network}
\newacronym{dl}{DL}{Deep Learning}
\newacronym{shm}{SHM}{Structural Health Monitoring}
\newacronym{svd}{SVD}{Singular Value Decomposition}
\newacronym{mri}{MRI}{Magnetic Resonance Imaging}
\newacronym{ir}{IR}{Infrared}
\newacronym{pt}{PT}{Pulsed Thermography}
\newacronym{tndt}{TNDE}{Thermographic NDE}
\newacronym{ppt}{PPT}{Pulsed Phase Thermography}
\newacronym{mt}{MT}{Modulation Thermography}
\newacronym{irt}{IRT}{Infrared Thermography}
\newacronym{pct}{PCT}{Principal Component Thermography}
\newacronym{sngan}{SNGAN}{Spectral Normalized GAN}
\newacronym{tsr}{TSR}{Thermographic Signal Reconstruction}
\newacronym{stft}{STFT}{Short-Time Fourier Transform}
\newacronym{kpca}{KPCA}{Kernel Principal Component Analysis}
\newacronym{dwt}{DWT}{Discrete Wavelet Transform}
\newacronym{abc}{ABC}{Artificial Bee Colony}
\newacronym{lstm}{LSTM}{Long Short-Term Memory}
\newacronym{ewt}{EWT}{Empirical Wavelet Transform}
\newacronym{emd}{EMD}{Empirical Mode Decomposition}
\newacronym{vmd}{VMD}{Variational Mode Decomposition}
\newacronym{gru}{GRU}{Gated Recurrent Unit}
\newacronym{roi}{ROI}{Region of Interest}
\newacronym{asrg}{ASRG}{Automatic Seeded Region Growing}
\newacronym{asbi}{ASBI}{Adaptive Spectral Band Integration}
\newacronym{gpct}{GPCT}{SNGAN-based generative principal component thermography}
\newacronym{fcnn}{FCN}{Fully Convolutional Neural Network}
\newacronym{rpn}{RPN}{Region Proposal Network}
\newacronym{ssd}{SSD}{Single Shot MultiBox Detector}
\newacronym{ssim}{SSIM}{Structural Similarity Index}
\newacronym{ihcp}{IHCP}{Inverse Heat Conduction Problem}

%% file: tikz_tables/tikzstyles.tex
\tikzstyle{line} = [draw, thick]
\tikzstyle{line1} = [draw, thick, ->]
\tikzstyle{line3} = [draw, gray, thick, dashed, ->]
\tikzstyle{line4} = [draw, blue, very thick, dashed, -]
\tikzstyle{largebox} = [rectangle, draw, line width=2pt, blue!50, 
text width = 2.5cm,  minimum height = 3.5cm
]
\tikzstyle{block1} = [rectangle, draw,  
text width=2.3cm, text centered, minimum height = 0.5cm
]
\tikzstyle{block2} = [rectangle, draw,  
text width=0.4cm, text centered, minimum height = 4cm
]
\tikzstyle{block3} = [rectangle, draw,  
text width=2.2cm, text centered, minimum height = 0.6cm
]
\tikzstyle{block4} = [rectangle, draw, fill = orange!12, 
text width=2.2cm, text centered, minimum height = 0.6cm
]

\newcommand{\ColOperation}{Apricot!35}
\newcommand{\ColEnhance}{Emerald!15}
\newcommand{\ColDecision}{CornflowerBlue!25}
\newcommand{\ColSynth}{CarnationPink!25}

%% file: sections/abstract.tex
\begin{abstract}
    The digital transformation is fundamentally changing our industries, affecting planning, execution as well as monitoring of production processes in a wide range of application fields. With product line-ups becoming more and more versatile and diverse, the necessary inspection and monitoring sparks significant novel requirements on the corresponding \gls{ndt} systems. The establishment of increasingly powerful approaches to incorporate \gls{ai} may provide just the needed innovation to solve some of these challenges.

    In this paper we provide a comprehensive survey about the usage of \gls{ai} methods in \gls{ndt} in light of the recent innovations towards \gls{ndt} 4.0. Since we cannot discuss each \gls{ndt} modality in one paper, we limit our attention to magnetic methods, ultrasound, thermography, as well as optical inspection. In addition to reviewing recent \gls{ai} developments in each field, we draw common connections by pointing out \gls{ndt}-related tasks that have a common underlying mathematical problem and categorizing the state of the art according to the corresponding sub-tasks. In so doing, interdisciplinary connections are drawn that provide a more complete overall picture.

\end{abstract}

%% file: sections/introduction.tex
\section{Introduction} \label{section_introduction}
\gls{nde} is one of the key ingredients for ensuring safety, reliability, and efficiency of products as well as infrastructure we use and depend on everyday. In light of challenges such as increasing technological complexity of our industries, aging infrastructure, or the general need to reduce resource consumption, its significance for industry and society is expected to grow dramatically in the near future. Like industry itself, the field of \gls{ndt} has undergone several transformations, adopting innovations in sensor technology, networking and computing power and more recently, \gls{ai} methods to cope with the requirements of increasingly complex industries. At the intersection between \gls{ndt} and Industry 4.0 technologies, the field of "NDE 4.0" has emerged as a confluence of Industry 4.0 digital technologies, physical inspection methods, and business models. It embraces the digital transformation, including digital twins of products and inspection systems, \gls{iiot} technologies for seamless networked integration, semantic interoperability through standardized interfaces, common data formats and appropriate meta descriptions, heterogeneous compute power through cloud and edge technologies, and other recent developments. By many, NDE 4.0 is also seen as a potential step to increase automation of current \gls{ndt} systems, as its growing complexity allows it to handle increasingly sophisticated tasks autonomously. 

The present paper provides a survey on the state of the art in the usage of \gls{ai} methods for \gls{ndt}, in light of the promises of the NDE 4.0 developments. We discuss to what extent certain levels of automation can now be handled through AI-powered \gls{ndt} approaches, which advances have been demonstrated in that regard recently, and which limitations we are still facing. It is illustrative to categorize the degree of NDE automation  according to \gls{ndt} automation levels as suggested in \citep{cantero2022deep}, which we later decompose into specific tasks, as well as the common \gls{ndt} operator levels according to DIN EN ISO 9712 \citep{iso9712}. A level 1 operator is mostly responsible for conducting measurements under the supervision of a higher level personnel and thus, the tasks are easier to automate. On the other hand, the tasks of a level 2 operator (result interpretation and decision making) or even a level 3 operator (modality selection, inspection design and full responsibility of the entire chain) require much higher degree of sophistication.

Since the entire discussion on \gls{ndt} automation is highly dependent on the considered modality, we treat a selection of \gls{ndt} modalities in different sections of this paper. As the entire field of \gls{ndt} is too wide to provide a thorough review in just one paper, we limit our scope to selected, representative modalities: magnetic methods (eddy current testing and micromagnetic methods), Ultrasound, thermography, and at the end also touching on optical inspection techniques.

As a second contribution, the paper attempts to provide connections between the AI-powered \gls{ndt} tasks and their underlying mathematical problems that are often similar. The common patterns help seeing intersections between related fields that can benefit from interdisciplinary work and give rise to powerful cross-domain AI approaches.

The paper is organized as follows: \cref{section_3ma} reviews magnetic methods, Ultrasound is treated in \cref{section_ultrasound}, in \cref{section_thermography} we discuss thermography, and \cref{opt._insp.} we consider optical inspection. The common patterns between AI-powered \gls{ndt} tasks are elaborated on in \cref{sec:overview} before drawing conclusions in \cref{section_summary}.

Throughout this work, quantities in bold lowercase $\bm{v}$ are vectors, while those in bold uppercase $\bm{M}$ are matrices. Scalar elements of vectors and matrices are denoted as $\bm{v}(i)$ and $\bm{M}(i,j)$, respectively. The symbol $( \cdot )\herm$ represents the Hermitian or complex conjugate transpose, and $( \cdot )^\dag$ is the pseudo-inverse. Calligraphic symbols $\mathcal{S}$ denote sets. Further symbols are introduced throughout the work as needed. 

%% file: sections/3ma.tex
\section{Magnetic methods} \label{section_3ma}
Several \gls{nde} methods rely on the magnetic interaction of a probe with the test specimen. These include, amongst others, eddy current testing of electrically conductive materials.
During testing, an alternating magnetic field is generated by a coil, which induces eddy currents in the material to be tested. The eddy current density then generates the secondary magnetic field which is measured with a sensor.  Typically, the measurement sensor also contains the excitation coil, enabling to infer the eddy current density.
The measured parameters are the amplitude and the phase shift to the excitation signal.
In the case of ferromagnetic materials, some magnetic properties, which correlate with the mechanical properties of the material, can also be detected with other techniques. 
For example, the Barkhausen effect, which is caused by the movement of the magnetic domain walls of the sample upon magnetic excitation, can be analyzed. 
A combination of those two methods together with the analysis of the harmonics of the tangential magnetic field and the incremental permeability \citep{DIN_1324}{\iffalse is given by the \fi} is possible, and among such mixed modalities we highlight the \gls{3ma} II system. 
The 3MA-X8 system is a further development, focusing on a more robust and less complex device and sensor design \citep{Sarg22}. This system also combines eddy current analysis, incremental permeability analysis and harmonic analysis, omitting the Barkhausen noise analysis \citep{Sarg22}. These methods find wide application in the field of material characterization as well as defect detection. 

\subsection{Eddy current testing} \label{sec:eddy_current}

Generally, eddy current probe impedance is recorded as a function of probe position and/or excitation frequency, which is used for detecting flaws such as cracks in materials. Inferring those parameters from the impedance measurements is an inverse problem and widely recognized as a complex theoretical problem. Already in 1997 Rekanos et al. \citep{RTP97} suggested the use of \glspl{nn} to tackle the inversion problem. Since then, a number of studies have focused on the detection and evaluation of flaws with the help of \glspl{ann}. Of particular interest here are defects such as cracks, which could paralyse operations and endanger safety. 

Thus, Helifa et al. \citep{HFL16}, Harzallah et al. \citep{HRC18}, Demachi et al. \citep{DHP19} and Bernieri et al. \citep{BFL08} dealt with the detection and assessment of cracks. While Helifa et al. and Harzallah et al. made use of \gls{ai} supported inversion for the detection and characterization of surface cracks, Bernieri et al. discussed crack shape reconstruction and compared the results using an \gls{ann} and \gls{svr} and Demachi et al. studied the estimation of crack depths in non-magnetic material using a \gls{cnn} with two convolution layers, two pooling layers and a fully connected layer. They were able to estimate the crack depth within an error of \SI{0.05}{mm}. For their research, Helifa et al. used a \gls{mlp} with 21 input neurons corresponding to 21 different sensor positions, a hidden layer with 80 neurons and hyperbolic tangent activation function and an output layer with a linear activation function. The network was trained on simulated data and achieved relative errors of \SI{2.5}{\%} for the depth and \SI{2}{\%} for the length of the cracks. Harzallah et al. took a similar approach and used a \gls{mlp} with 1 input layer, 1 hidden layer and 1 output layer and trained it on simulated signals. They stopped the training after 30 epochs at an \gls{mse} of about \SI{5e-7}{\milli\metre}$^{2}$ for the train and validation data, while the test error was about \SI{1e-2}{\milli\metre}$^{2}$ for length and depth of the crack. 
On the other hand, Bernieri et al. discussed crack shape reconstruction and compared the results using an \gls{ann} and \gls{svr}. The database for the training was constructed by simulation. A \SI{200}{\milli\meter} × \SI{200}{mm} × \SI{2}{mm} aluminum plate was simulated, where the defect region was divided into a regular grid of 20 × 4 elements, imposing defect depth $z_0$ and height $a$ steps of \SI{0.5}{mm} and length $l$ step of \SI{1}{mm}. This way, a data set of 200 simulated magnetic field maps was obtained. They found that the \gls{svr} outperformed the three layered \gls{mlp} in terms of \gls{mae} as shown in \cref{tab:tab_A}.

\begin{table}[ht]
\caption{Crack shape reconstruction results of Bernieri et al. \citep{BFL08}}
\label{tab:tab_A}
\begin{center}
\begin{tabular}{ m{1.6cm} m{1.7cm} m{1.7cm} m{1.6cm} } 
 \hline
            & MAE $l$   & MAE $a$   & MAE $z_0$ \\
 \hline
 \hline
    SVR     & 0.043     & 0.12      & 0.12 \\
    ANN     & 0.14      & 0.24      & 0.24 \\
 \hline
\end{tabular}
\end{center}
\end{table}

If we move into the area of nuclear reactors, not only cracks but even minor flaws are of great interest and must be reliably recognised for safe operation. Thus, Song et al. \citep{SS99} as well as Yusa et al. \citep{YCC02} have dealt with the investigation of flaws in steam reactor pipes. In order to train a \gls{nn} on an synthetic dataset generated by simulation to determine the type and size parameters of flaws in steam generator tubes, Song et al. \citep{SS99} built a synthetic database with 400 ECT signals generated from 200 axisymmetric machined grooves in four types with two test frequencies per flaw. They used a probabilistic \gls{nn} for the flaw classification and a back propagation network for the flaw sizing. This resulted in an overall accuracy of \SI{91}{\%} in the determination of the flaw type while the correlation coefficients between actual and estimated size parameters exceeded 0.97. For the reconstruction of natural cracks that occurred in steam generator tubes, Yusa et al. \citep{YCC02} modeled cracks as a pixel map and simulated eddy current signals from that. \glspl{nn} were trained using a database that contained 400 cases of simulated data. Subsequently, genuine natural cracks that were found in a steam generator tube of a nuclear power plant were measured and reconstructed using the trained network, demonstrating good agreement with the results of the destructive inspection carried out afterwards.
Wrzuszczak \citep{WW02} detected flaws in conducting layers of aluminum and copper as well as in ferrous tubes by using \glspl{nn}. They tried \gls{mlp}s and \gls{rbf} networks and showed, that the \gls{rbf} outperformed the \gls{mlp} on this task, resulting in a lower training time.

Many other \gls{ect} tasks also can be improved by the aid of \gls{ai} methods. For instance, Glorieux et al. \citep{GMB99} studied the electrical conductivity profiles using \gls{nn} inversion of multi-frequency eddy current data. A \gls{mlp} consisting of three sigmoid input neurons and one linear output neuron was trained on 2000 simulated conductivity profiles which was obtained using piecewise constant functions (20 segments). Afterwards the network was tested on real data in form of artificially stacked samples and showed good results. On the other hand, Rao et al. \citep{RRJ02} used \gls{nn}s to test austenitic stainless steel welds in order to detect and characterize longitudinal and transverse surface-breaking notches. On the basis of multi-frequency eddy current data they evaluated the depth of the surface-breaking notches in the welds with a maximum deviation of \SI{0.08}{mm}. Kuzmin et al. \citep{KGP19} applied \gls{nn}s for recognizing rail structural elements in magnetic and eddy current defectograms. They investigated three classes of structural elements of a railroad track: (1) a bolted joint with a straight or beveled rail connection, (2) a flash butt rail weld, and (3) an aluminothermic rail weld. Patterns that couldn't be assigned to these three classes were considered as defects and attributed to a separate fourth class. The used \gls{nn} contained 780 input neurons, a hidden layer with 800 neurons and four output neurons. The training set contained 16287 real and 8323 generated samples of structural elements of three classes and 11417 real samples of conditional defects. The validation set contained 4098 generated samples of structural elements of three classes and 1903 real samples of conditional defects, whereas the test set consisted of 4098 and 1902 samples of the same kinds, respectively. The achieved accuracy was \SI{99.83}{\%} on training data, \SI{99.45}{\%} on validation data and \SI{99.23}{\%} on test data.
In order to identify broken wires within a wire rope, Cao et al. \citep{CLH11} used a \gls{rbf} network (3 input neurons, 2 output neurons, one hidden layer) and were able to not only detect if there was a broken wire, but also estimate how many wires were broken. Further Ali et al. \citep{AAR17} gave a review on system development in \gls{ect} and technique for defect classification and characterization as well as an overview of AI methods used in defect measuring, while AbdAlla et al. \citep{ASF18} gave a review of challenges in improving the performance of \gls{ect} as well as of \gls{ai} methods used in defect measuring. 

\subsection{Barkhausen effect} \label{sec:berkhausen}
The Barkhausen effect was discovered in 1919 by the German physicist Heinrich Barkhausen. Barkhausen noise consists of discontinuous changes in the magnetization of ferromagnetic materials in a changing external magnetic field. This is explained by the fact that the shifting Bloch walls are temporarily "trapped" by the lattice perturbations and only break free from their anchoring abruptly when the driving force is increased (Barkhausen event). A further increase of the field strength causes the magnetization in the domains to rotate in the direction of the external magnetic field. When applying an external alternating magnetic field to a ferromagnetic specimen, the resulting Barkhausen noise can be recorded with a Hall probe. Since the Barkhausen events are tightly connected to the lattice defects in the material, the characteristics of the recorded Barkhausen noise correlate with the mechanical properties of the material, which are also linked to lattice defects. 

Consequently this enables the evaluation of magnetic properties of the material under investigation, like done by Maciusowicz et al. \citep{MP22}, but also the mechanical properties of the material, like Wang et al. \citep{WZZ13} and Zhang et al. \citep{ZLL19} show. Consequently, this enables the evaluation of magnetic properties of the material under investigation as well as the mechanical properties of the material. Aiming to identify the magnetic porperties of the test materials, Maciusowicz et al. \citep{MP22} evaluated the grade and the magnetic directions of conventional and high grain oriented electrical sheets subjected to selected surface engineering methods. 
Several \gls{ml} techniques (decision trees, discriminant analysis, \gls{svm}, naïve Bayes, \gls{knn}, ensemble classifiers and \gls{ann}) are compared. The achieved results generally indicated an approximately \SI{10}{\%} advantage of the deep learning model over the classical ones in terms of accuracy in each of the considered cases.
For determining the mechanical properties of the test objects, Wang et al. applied a \gls{nn} for the steel stress detection based on Barkhausen noise in order to achieve robust results against temperature fluctuations, which have an impact on the stress as well as on the received signal. Zhang et al. predicted yield strength, ultimate tensile strength and elongation of cold rolled strips by the use of a fully connected \gls{nn} with two hidden layers (five input nodes, three output nodes, five nodes in first hidden layer and six nodes in second hidden layer). In the test results the largest relative errors of yield strength, tensile strength and elongation are \SI{5.66}{\%}, \SI{2.99}{\%} and \SI{3.62}{\%}, respectively.
\subsection{3MA II and further combinations of methods} \label{sec:3ma2}

By combining several methods into one framework, the robustness, reliability and precision with regard to the determination of mechanical properties can be increased. An example of such a combination of methods is given by the 3MA II device, which not only combines eddy current and Barkhausen noise measurements, but further also measures the harmonic distortion of a pure sinusoidal exitation as well as the incremental permeability of the material. 
Thus 3MA or similar devices are limited to ferromagnetic materials, but can be used in a wide range of applications, including but not limited to the estimation of mechanical properties, as Jedamski et al. \citep{JE20}, Wang et al. \citep{WHL22}, Xing et al. \citep{XWN23} as well as Sheng et al. \citep{SW23} have done in their work, whereby Sheng et al. further combined the 3MA II with \gls{emat}.

In their work, Jedamski et al. \citep{JE20} determined the hardness and the case hardening depth using linear regression as well as \glspl{nn} and compared the results. Their sample set consists of 54 discs made from one batch of steel AISI 4820 (DIN 18CrNiMo7-6) with a diameter of \SI{68}{mm} and a thickness of \SI{20}{mm}. The samples were gas carburized and oil quenched in 27 variations. The best results achieved are shown in \cref{tab:tab_B}.

\begin{table}[ht]
\caption{Results of the estimation of hardness and carburization depth from Jedamski et al. \citep{JE20}, comparing the linear regression and the neural network approach}
\label{tab:tab_B}
\begin{center}
\begin{tabular}{ m{1.15cm} m{2.3cm} m{1.5cm} m{1.5cm} } 
 \hline
                                            &                   & RMSE Hardness (HV1)   & RMSE Carburization Depth (mm) \\
 \hline
 \hline
    \multirow{2}{1.15cm}{Training score}    & linear regression & 12.5                  & 0.074 \\\cline{2-4}
                                            & neural network    & 7.1                   & 0.03 \\
 \hline
    \multirow{2}{1.15cm}{Test score}        & linear regression & 17.9                  & 0.13 \\\cline{2-4}
                                            & neural network    & 16.4                  & 0.103 \\
 \hline
\end{tabular}
\end{center}
\end{table}


Both Wang et al. \citep{WHL22} and Xing et al. \citep{XWN23}, predicted the surface hardness in ferromagnetic Materials by using a device quite similar to 3MA II. 
Wang et al. \citep{WHL22} studied the prediction of surface hardness in carbon steels, referred as Cr12MoV steel, 45 steel, and martensitic stainless steel of 3Cr13 in Chinese standards. For each material, a total of 27 specimens  (\SI{200}{mm} × \SI{60}{mm} × \SI{3}{mm}) were cut from a steel plate along its rolling direction. Afterwards, quenching, followed by tempering, was performed. The \gls{nn} used to predict the surface hardness consisted of 31 input neurons for selected magnetic measurands, one hidden layer with 50 neurons and one output neuron. The achieved \gls{rmse} of the predictions of surface hardness was \SI{7.52}{HV30} for the Cr12MoV steel, \SI{4.27}{HV30} for the 3Cr13 steel and \SI{7.44}{HV30} for the 45 steel. Xing et al. \citep{XWN23} on the other hand, evaluated the surface hardness only on Cr12MoV steel. For that purpose, a batch of 8 plates with dimensions of \SI{250}{mm} x \SI{60}{mm} x \SI{3}{mm} were cut from the cold work die steel plates and were quenched. Afterwards, hardness (HV30) was tested on 3 positions per plate. A simple feed forward network with an input layer for the micromagnetic measurands, two hidden layers with 10 nodes and an output layer with one node for the hardness was used. The \gls{mae} and \gls{sd} in surface hardness prediction were approximately \SI{1.02}{HV} and \SI{1.04}{HV}, respectively.


In order to to predict the \gls{rm} and surface hardness (HA) in pipeline steel, Sheng et al. \citep{SW23} further combined 3MA II with \gls{emat}. Training and validation data were acquired on seventeen pipeline steel samples (\SI{300}{mm} × \SI{750}{mm}, thickness > \SI{12}{mm}) of various lots and grades. For the test data an X65 pipeline steel plate with a size of \SI{4}{m} × \SI{12}{m} with a tensile strength of \SI{670}{MPa} was prepared. The tensile strength was assumed to be constant over the whole plate, hardness was tested at 65 points distributed over the plate. The prediction results show that, compared to \gls{mlp}, \gls{rf}, and \gls{dnn}, the approach based on the \gls{xgboost} was found to yield the highest prediction accuracy in terms of \gls{rmse} and \gls{sd} for the test data set, as shown in \cref{tab:tab_C}. 

\begin{table}[ht]
\caption{Comparision of the results from different approaches to estimate tensile strength and hardness from Sheng et al. \citep{SW23} }
\label{tab:tab_C}
\begin{center}
\begin{tabular}{ m{2.3cm} m{2.15cm}  m{2.5cm} } 
 \hline
            & \gls{rm}        & HA \\
            & \gls{sd} (MPa)  & \gls{rmse} (HV1) \\
 \hline
 \hline
 XGBoost    & 17.69                          & 11.85 \\
 MLP        & 31.9                           & 25.45 \\
 RF         & 19.47                          & 15.49 \\
 DNN        & 18.47                          & 14.22 \\
 \hline
\end{tabular}
\end{center}
\end{table}

Another example of \gls{ai} aided estimation of mechanical properties based on the combination of several techniques is given by Grönroos et al. \citep{GRK21}, who estimated the embrittlement of a nuclear reactor pressure vessel, which is measured as the \gls{dbtt}. For that purpose, they combined 3MA II with Micromagnetic Inductive Response and Barkhausen Emission Method, Magnetic Adaptive Testing, Direct Current Reversal Potential Drop, Thermoelectric Power Measuring Method and \gls{ut} to collect data on standard Charpy V-notch samples made out of six different steel alloys that had been treated in different irradiation conditions. Their data set consists of 29 nondestructively measured parameters and ductile-brittle transition temperature data for 157 samples. Several machine learning methods were compared: linear \gls{hlr}, \gls{svr} and an \gls{ann} consisting of 19 input neurons, one output neuron, 2 hidden layers with 16 and 8 neurons respectively and two dropout layers in between. The results in terms of \gls{rmse}, \gls{mae} and \gls{r2} are shown in \cref{tab:tab_D}.

\begin{table}[ht]
\caption{Comparision of the results from different approaches to estimate the \gls{dbtt} from Grönroos et al. \citep{GRK21} }
\label{tab:tab_D}
\begin{center}
\begin{tabular}{ m{2cm} c c c c } 
 \hline
						              &	                      & \gls{hlr}   & \gls{svr}   & \gls{ann} \\ 
 \hline
 \hline
 10-fold cross-validation score     & \gls{mae} (°C)              & 13.67 & 13.58 & 12.7 \\ 
 \hline
 \multirow{3}{4em}{Training score}  & \gls{mae} (°C)              & 9.32  & 8.68  & 9.21 \\ 
                                    & \gls{rmse} (°C)             & 14.49 & 13.63 & 12.57 \\
                                    & \gls{r2}  & 0.97  & 0.97  & 0.98 \\
 \hline
                         
  \multirow{3}{4em}{Test score}     & \gls{mae} (°C)              & 17.1  & 17.77 & 16.04 \\
                                    & \gls{rmse} (°C)             & 18.67 & 20.4  & 22.08 \\
                                    & \gls{r2}  & 0.95  & 0.94  & 0.93 \\
 \hline
\end{tabular}
\end{center}
\end{table}

\subsection{3MA-X8} \label{sec:3max8}

In comparison to the \gls{3ma} II, omitting the Hall sensor, as well as the Barkhausen noise and eddy current coils simplified the probe to a voltage-controlled electromagnet \citep{Sarg22}. By exciting the electromagnet with a low- and high frequency component and measuring the voltage and current of the probe, features of the three methods of measurement, eddy current analysis, incremental permeability analysis and harmonic analysis, are extracted \citep{Sarg22}. This simplification leads to a more robust system better suited for higher measurement speeds and sensor arrays \citep{Sarg22}. The implementation of \gls{ml}-Methods considerably enhanced precision and robustness of the \gls{3ma}-X8 system even further. New approaches improved the application fields of process optimization in steel mills and reactor safety research in particular \citep{Sarg21,Cy21}.


Youssef et al. \citep{Sarg19dg} describes initial studies on improving the accuracy by extending the feature space. In the feature space extension, the available time signals are analyzed in more detail by using Fourier transform, a qualitative hysteresis reconstruction and a multidimensional curve tracing approach, where mathematical characteristics such as high points, zero points and axis intercepts are included as features. The underlying data was acquired in a tensile test on three different normalized materials, 100Cr6, C45 and S235 where each material was stressed in increments of $\SI{5}{\%}$  up to $\SI{50}{\%}$ of its \gls{re}. To classify the stress levels independently of the material, a \gls{lda} reduces the feature space down to two dimensions before a \gls{knn}-Classifier with \textit{k} = 10 predicts the class. Regarding the classifiability a \gls{kfcv} with a randomized train-test-split and a \textit{k} = 10 was used to determine the accuracy before and after feature space extension. Considering the quantifiability of the stress levels, a class based validation assigns every second stress level to the test data set. Both, the classifiability and the quantifiability were improved significantly, as shown in \cref{tab:tab_Sarg19dg} below using the accuracy. To further describe the quantifiability, a \gls{mlr} predicts the stress level also based on a class based validation assigns every second stress level to the test data set. The quantifiability was also improved significantly, as shown in \cref{tab:tab_Sarg19dg}.

%

\begin{table}[ht]
\caption{Classifiability and quantifiability comparison of the \gls{3ma}-X8 feature spaces on tensile test data \citep{Sarg19dg}}
\label{tab:tab_Sarg19dg}
\begin{center}
\begin{tabular}{ m{2.1cm} m{1cm} m{1cm} m{1cm}  m{1cm} } 
 \hline
  &\multicolumn{2}{c}{ \gls{lda} \& \gls{knn} }  \\    
  &\multicolumn{2}{c}{ accuracy (\%) }  & \multicolumn{2}{c}{\gls{mlr}}   \\
  Feature space        & Random \gls{kfcv}    & Class based \newline validation  &  \gls{rmse} \newline  (\% \gls{re})   & \gls{r2} \newline  (\%) \\
 \hline
\hline
  21-dimensional       & 64.05  & 63.08 & 4         & 93.6    \\
  Extended             & 95.14  & 94.21 & 1.96      & 98.46    \\
 \hline
\end{tabular}
\end{center}
\end{table}


Youssef et al. \citep{Sarg19ama} further investigates different \gls{ml} methods based on the same experimental setup as in the previous study \citep{Sarg19dg}, while also comparing the performances of the feature spaces. Specifically, it was compared whether and with what effort the prediction quality of \gls{mlr} can be achieved under similar conditions with \gls{svr} and \gls{mlp}-Regression with a single hidden layer and linear activation functions, which are frequently used in regression problems. Under variation of the regularization parameter \textit{c} in the \gls{svr} and variation of the number of neurons in the hidden layer of \gls{mlp}-Regression, the best performing model was chosen. The train data \gls{rmse} and \gls{r2} of these models are shown in \cref{tab:tab_Sarg19ama}. The extended feature space has performed significantly better overall. However, neither the \gls{svr} nor the \gls{mlp}-Regression produced significant improvements under the conditions explained.

\begin{table}[ht]
\caption{Different \gls{ml} methods of the \gls{3ma}-X8 feature space on tensile test data \citep{Sarg19ama}}
\label{tab:tab_Sarg19ama}
\begin{center}
\begin{tabular}{ m{1.3cm} m{2.2cm} m{0.75cm} m{1cm} m{1cm} } 
 \hline
     Model                                      & Feature space                         &                   & \gls{rmse} (\% \gls{re})  & \gls{r2} \newline (\%) \\
 \hline
 \hline
    \multirow{2}{1.15cm}{\gls{mlr}}             & \multirow{2}{2.2cm}{21-dimensional}   & Train             & 7.99                      & 78.07 \\
                                                &                                       & Test              & 7.83                      & 69.35 \\
\cline{3-5} 
    \multirow{2}{1.15cm}{\gls{mlr}}             & \multirow{2}{2.2cm}{Extended}         & Train             & 2.3                       & 98.19 \\
                                                &                                       & Test              & 4.99                      & 87.54 \\
\cline{3-5} 
    \multirow{2}{1.15cm}{\gls{svr}}             & \multirow{2}{2.2cm}{21-dimensional}   & Train             & 8.32                      & 69.85 \\
                                                &                                       & Test              & 7.62                      & 66.69 \\
\cline{3-5} 
    \multirow{2}{1.15cm}{\gls{svr}}             & \multirow{2}{2.2cm}{Extended}         & Train             & 4.11                      & 93.96 \\
                                                &                                       & Test              & 4.49                      & 90.33 \\
\cline{3-5} 
    \multirow{2}{1.15cm}{\gls{mlp}-Regression}  & \multirow{2}{2.2cm}{21-dimensional}   & Train             & 10.33                     & 63.42 \\
                                                &                                       & Test              & 9.48                      & 55.08 \\
\cline{3-5} 
    \multirow{2}{1.15cm}{\gls{mlp}-Regression}  & \multirow{2}{2.2cm}{Extended}         & Train             & 5.54                      & 89.49 \\
                                                &                                       & Test              & 5.69                      & 83.82 \\
 \hline
\end{tabular}
\end{center}
\end{table}

Youssef et al. \citep{Sarg21} further published the application of \gls{ml} methods with the \gls{3ma}-X8 in process optimization in steel mills, proving the successful application of the prior investigated methodologies \citep{Sarg19ama,Sarg19dg} in an industrially relevant environment. The focus was on the detection of local hardness inhomogeneities in heavy plates for the pipeline industry, where the task was to detect hardness differences of +/- 30 HV 10 at a minimum diameter of $\SI{10}{\mm}$. Supervised \gls{ml} methods sufficiently suppressed various interferences coming from residual fields and scale layers. Unsupervised \gls{ml} algorithms suppressed the lift-off influence based on a series of measurements with a sensor lift-off variation of $\SI{0}{\mm}$ to $\SI{3}{\mm}$ on sample heavy plate segments. Based on these findings, the product "PLAMAT-MM-32M" was developed for the use in steel mills. %

\begin{figure}[ht]
\centering
\includegraphics[width=0.45\textwidth]{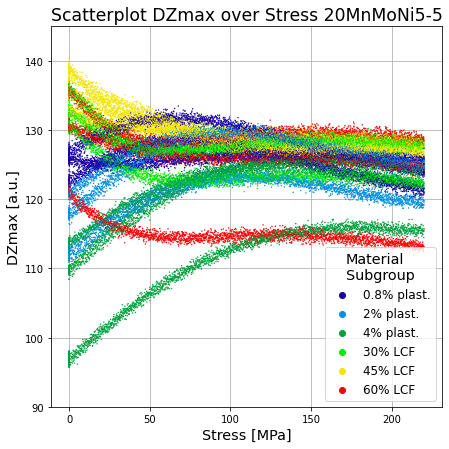}
\caption{\centering 20MnMoNi5-5 stress-dependent scatterplot of feature DZmax (maximum value of incremental permeability) which is not sufficient to differentiate the material variants considered in \citep{Cy21} 
}
\label{fig:DZmax}
\end{figure}

\begin{figure}[ht]
\centering
\includegraphics[width=7.7cm]{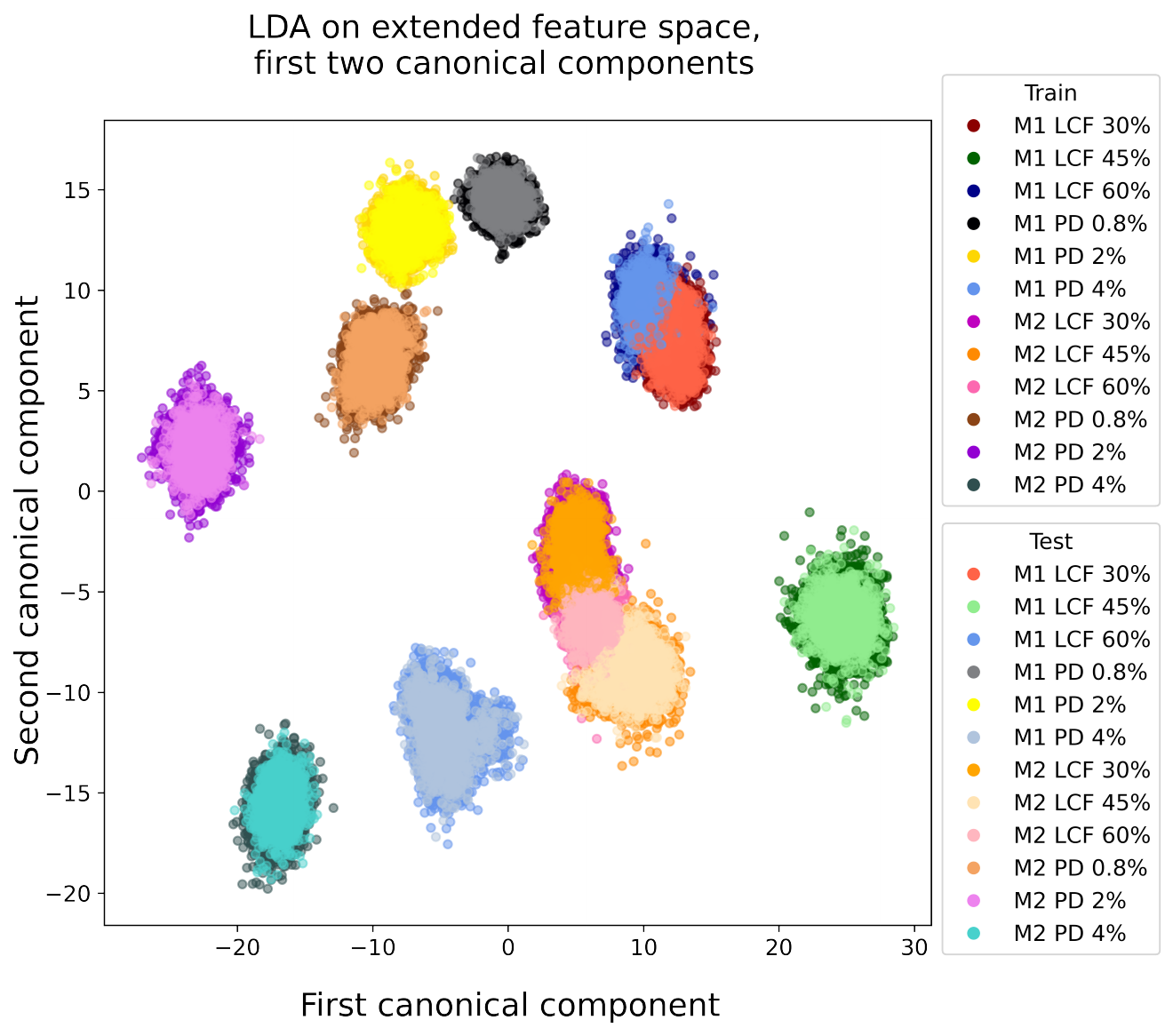}
\caption{\centering First two canonical components transformed by LDA of the extended feature space \citep{Cy21}, demonstrating a clearer separation of individual material variants compared to \cref{fig:DZmax}. "M1" refers to the material 20MnMoNi5-5 and "M2" refers to the material 22NiMoCr3-7} %

\label{fig:Cy21Scat}
\end{figure}

Zimmer et al. \citep{Cy21} described the application on \gls{rpv}-steel, where the objective is to differentiate between stress and microstructural influences. The materials used are 20MnMoNi5-5 and 22NiMoCr3-7 with further variation levels introduced by \gls{lcf} tests and tensile tests which leads to a more complex dataset in comparison to prior studies \citep{Sarg19ama,Sarg19dg}. Regarding the features of the dataset, the extended feature space was further extended by the raw signal samples as features themselves. To illustrate the ambiguities between microstructure and stress influences, \cref{fig:DZmax} shows a single feature of the 21-dimensional feature space. For stress-independent microstructure characterization, a train-test-split randomly divided the dataset into $\SI{80}{\%}$ train data and $\SI{20}{\%}$ test data. Based on this dataset, an \gls{lda} performed a dimensional reduction to four dimensions. The first two dimensions shown in \cref{fig:Cy21Scat} indicate a much clearer separation of individual material variants illustrated by the the point clouds. Subsequently, a \gls{knn}-Classifier with \textit{k} = 5 neighbors and uniform weighting was trained. The extended feature space significantly improves the model with $\SI{0.77}{\%}$ misclassified data points of the test data compared to $\SI{3.89}{\%}$ of the 21-dimensional feature space. For microstructure-independent stress determination, a \gls{mlr} predicted the stress in MPa. For the train-test split, the dataset was divided into seven segments along the target variable, alternating between train and test data to detect overfitting and overgeneralization. 
A hierarchical approach was performed, dividing the dataset in different subgroups prior to fitting the \gls{mlr}. This approach was able to improve the models accuracy significantly in comparison to the global models as shown in \cref{tab:tab_Cy21}.

\begin{table}[ht]
\caption{Hierarchical modeling of the \gls{3ma}-X8 feature space on \gls{rpv}-steel data \citep{Cy21}}
\label{tab:tab_Cy21}
\begin{center}
\begin{tabular}{ m{2.4cm} m{2.25cm} m{1cm} m{1cm} }
\hline
   Feature Space                        & Subgroup         & Train \gls{rmse} (MPa) & Test \gls{rmse} (MPa) \\
\hline
\hline
    21-dimensional                      & Global           & 43.91                  & 40.10 \\               
    Extended                            & Global           & 13.98                  & 14.66 \\
    Extended                            & Tensile tests    & 9.94                   & 11.01\\              
    Extended                            & \gls{lcf} tests  & 9.36                   & 10.87\\              
    Extended                            & 20MnMoNi5-5      & 8.14                   & 8.93 \\              
    Extended                            & 22NiMoCr3-7      & 10.91                  & 13.95\\
\hline
\end{tabular}
\end{center}
\end{table}

Youssef et al. \citep{Sarg22} systematically examined the previous approaches and combined them into a global methodology. In order to map nonlinear relationships between the feature space and the target variable, a polynomial extension linearizes the feature space iteratively, until the model quality reaches an optimum. This systematic approach was applied on the \gls{rpv}-steel materials and experimental setup of the prior paper \citep{Cy21}. To determine the stress in \% \gls{re}, a \gls{mlr} was conducted. Regarding the train-test split, a segment based approach was used, dividing the dataset into segments along the target variable alternating between train and test data. Before the main model was trained, the number of segments was determined by iteratively increasing the number of segments, until no significant improvement in the \gls{mlr} occurs. As a result, the train-test split divided the data into 21 train segments and 20 test segments. To show the improvement of the iterative polynomial extension, different iterations and feature spaces were examined as shown in \cref{tab:tab_Sarg22}. Comparing the eleventh polynomial grade of the 21-dimensional feature space to the fourth polynomial grade of the extended feature space it is clear, that in this example the use of the extended feature space yields a significantly better result with fewer iterations.

\begin{table}[ht]
\caption{Iterative polynomial extension of the \gls{3ma}-X8 feature space \citep{Sarg22}}
\label{tab:tab_Sarg22}
\begin{center}
\begin{tabular}{  m{2.2cm} m{1.2cm} m{0.85cm} m{1cm} m{1cm}  } 
 \hline
    Feature Space                        &  Polynomial Grade         &         & \gls{rmse} (\% \gls{re})       & \gls{r2} (\%) \\
 \hline
 \hline
    \multirow{2}{2.2cm}{21-dimensional}  & \multirow{2}{0.6cm}{1}    & Train   & 6.721                          & 81.0 \\
                                         &                           & Test    & 6.567                          & 78.2 \\
\cline{3-5} 
    \multirow{2}{2.2cm}{21-dimensional}  & \multirow{2}{0.6cm}{11}   & Train   & 3.014                          & 96.2 \\
                                         &                           & Test    & 2.863                          & 95.9 \\
\cline{3-5} 
    \multirow{2}{2.2cm}{Extended}        & \multirow{2}{0.6cm}{4}    & Train   & 0.830                          & 99.7 \\
                                         &                           & Test    & 0.974                          & 99.5 \\
 \hline
\end{tabular}
\end{center}
\end{table}

%% file: sections/ultrasound.tex
\section{Ultrasound} \label{section_ultrasound}


Ultrasound signals, {characterized by frequencies from 20 kHz to the gigahertz range and beyond human hearing, are pivotal in computational imaging, especially in \gls{ndt}. A key technique in this domain} is pulse-echo imaging, {which} involves the capture and utilization of scattered echoes arising from inhomogeneities within the measurement area.

In its simplest form, ultrasound data can be modeled via the wave equation \citep{verweij2014simulation}, so that the measured pressure field $p$ obeys
\begin{align} \label{eq_wave}
    \nabla^2 p - \frac{1}{c^2} \frac{\partial^2 p}{\partial t^2} = -s.
\end{align}
In \eqref{eq_wave}, $p = p(t, \bm{x})$ is a pressure field that depends on the time $t$ and position $\bm{x} \in \real^d$ for a $d$-dimensional space. The speed of sound $c = c(\bm{x})$ may in general vary over space due to changes in density and compressibility, and the term $s = s(t, \bm{x})$ represents sources in the medium. Alternatively, when working with a monochromatic field or after separation of variables, the frequency domain representation of the field obeys the Helmholtz equation, meaning
\begin{align} \label{eq_helm}
    \nabla^2 \tilde p + k^2 \tilde p = -\tilde s,
\end{align}
where quantities with a tilde $\tilde{(\cdot)}$ are frequency domain counterparts to the corresponding quantities without the hat. As a result, $\tilde p = \tilde p(\omega, \bm{x})$ and $\tilde s = \tilde s(\omega, \bm{x})$ depend on the angular frequency $\omega$, and all quantities including the wavenumber $k = k(\omega, \bm{x})$ can vary over space.

A common practice is to obtain approximate solutions to \cref{eq_wave,eq_helm} by discretizing them into systems of linear equations. In the time domain, the problem can be formulated as
\begin{align}
    \bm{p} = \bm{G \gamma},    
\end{align}
where $\bm{p}$ is obtained by discretizing $p$, e.g. along the temporal and spatial axes. The matrix $\bm{G}$ is referred to as the \emph{model} or \emph{measurement} matrix, and its formulation depends on the task. Similarly, the vector $\bm{\gamma}$ encodes the sought-after parameters, and its exact definition depends on the application.

Many ultrasound-based \gls{ndt} techniques revolve around estimating the speed of sound $c$, the wavenumber $k$, or the underlying physical parameters that determine them. Quantitative and qualitative estimates of these parameters are employed in detection, localization, imaging, classification, and more. Additionally, the amount of information about these parameters depends on the way in which the data is measured. As a result, the optimization of measurement parameters is another well-researched task. We discuss recent trends and open opportunities in the application of artificial intelligence to ultrasound \gls{ndt} tasks next.

\subsection{Ultrasound Imaging} \label{sec:us_ssr}


Conventional techniques like \gls{pwi} and \gls{fmc} transmit several incident waves from different angles and/or locations and collect the scattered echoes with one or more receivers \citep{le2016plane}. These traditional techniques mostly depend on augmenting the number of transmitted signals to enhance the accuracy and resolution of the reconstruction or classification.  

A mathematical model for the received signal in the ultrasound imaging can be defined in the form of matrix-vector multiplication as
\begin{align}
{\bm {\tilde p}}_{\mathrm{rec}}={\bm G}{\boldsymbol{\gamma}} + {\bm{\tilde \eta}}.
\end{align}
where ${\bm {\tilde p}}_{\mathrm{rec}}$ is the received or measured frequency domain signal vector carrying distinct samples, ${\boldsymbol{\gamma}}$ is the parameter of interest, and ${\bm{\tilde \eta}}$ represents complex additive noise. A common choice for the construction of ${\bm G}$ is to employ the first-order Born approximation and Green's function for a homogeneous, isotropic medium, as shown in \citep{schiffner2019random}. Following this model, the inverse scattering problem in which $\bm{\gamma}$ is estimated from the measurement data ${\bm{\tilde p}}_{\mathrm{rec}}$ can be reformulated as an optimization problem of the form
\begin{align} \label{inverse}
\hat{\bm{\gamma}} = \argmin_{\bm{\gamma}} \lambda R(\bm{\gamma}) + \left \Vert \tilde{\bm p}_{\mathrm{rec}} - \bm{G \gamma} \right \Vert_2^2,
\end{align}
where $R(\cdot)$ is a regularizer and $\lambda$ regulates its impact on the solution. 

One example for the solution of \cref{inverse} is the case where $\bm{\tilde p}_{\text{rec}}$ is represented sparsely by ${\bm{G}}$. In that case, $R(\cdot) = \Vert \cdot \Vert_1$ and $\bm{\gamma}$ can be estimated by harnessing a reduced amount of measurement data without reducing performance; in some cases, even by exploiting just a single acquisition from each element of the ultrasound transducer array with optimal excitation \citep{schiffner2019random, schiffner_america, cakiroglu, camsap23}. There are several conventional sparse data recovery algorithms, i.e. the \gls{ista} \citep{daubechies2004iterative} based on proximal gradient descent, its accelerated version \gls{fista} \citep{4959678} and the \gls{admm} \citep{boyd2011distributed} which splits the problem into simpler subproblems which separately focus on addressing the reconstruction fidelity and promoting sparsity through a penalty function. However, all of these algorithms work iteratively and come with a significant computational cost and time expense. Hence, novel methodologies based on neural networks have emerged that lower the reconstruction time without compromising the reconstruction accuracy. 

In the next sections, we explore different artificial intelligence approaches to the inverse problems involved in imaging. One approach focuses on designing conventional neural network architectures to tackle the inverse problem, while the other involves reconstruction based on the model of the data and well-known classical optimization algorithms. 

\subsubsection{Ultrasound Imaging with Conventional Deep Learning Architectures} \label{sec:ut_imaging_vanillaDL}

A self-supervised auto-encoder network {that minimizes} the \gls{mse} between $\bm{{p}}_{\mathrm{rec}}$ and $\hat{\bm{p}}_{\mathrm{rec}}$ is employed in \citep{zhang2020self}. {This network processes the projection of measurement data in the image domain}, i.e. the intermediate quantity ${\boldsymbol{\bar{\gamma}}} = \bm{G}\herm \bm{p}_{\mathrm{rec}}$ {which resembles} \gls{saft}. After the autoencoder network yields the reconstruction $\bm{\hat{\gamma}}$ as training output, its projection to the measurement domain is calculated this time as ${\bm{\hat{p}}}_{\mathrm{rec}} = \bm{G}{\boldsymbol{\hat{\gamma}}}$. The results show that the autoencoder gives a similar \gls{cnr} and the same axial \gls{fwhm} as conventional \gls{das} beamforming using 75 \glspl{pw}, as well as a lower \gls{fwhm} in lateral direction, in spite of exploiting only one \gls{pw}. {Similarly}, based on \gls{drl} \citep{he2016deep}, a method for removing the vanishing gradient problems in the neural networks is proposed in \citep{yao2019dr2}. {Firstly}, the measured signal is linearly mapped to the image domain using a fully connected network as ${\boldsymbol{\bar{\gamma}}} = \bm{W}\bm{p}_{\mathrm{rec}}$ and the elements of the matrix $\bm{W}$ are learned using the conventional \gls{sgd} training algorithm. After obtaining the initial estimation of the reconstruction data, a second algorithm employing \gls{drl} is used to obtain $\bm{\hat{\gamma}}$ from the initial estimate $\bm{\bar\gamma}$. The authors perform a comparison of \gls{psnr} with another residual sparse data reconstruction network called ReconNet \citep{7780424}. {The algorithm demonstrates improved \gls{psnr} and faster reconstruction than ReconNet.}

\begin{figure}[h]
\label{fig_unfolding}
\centering
\includegraphics[scale=0.5]{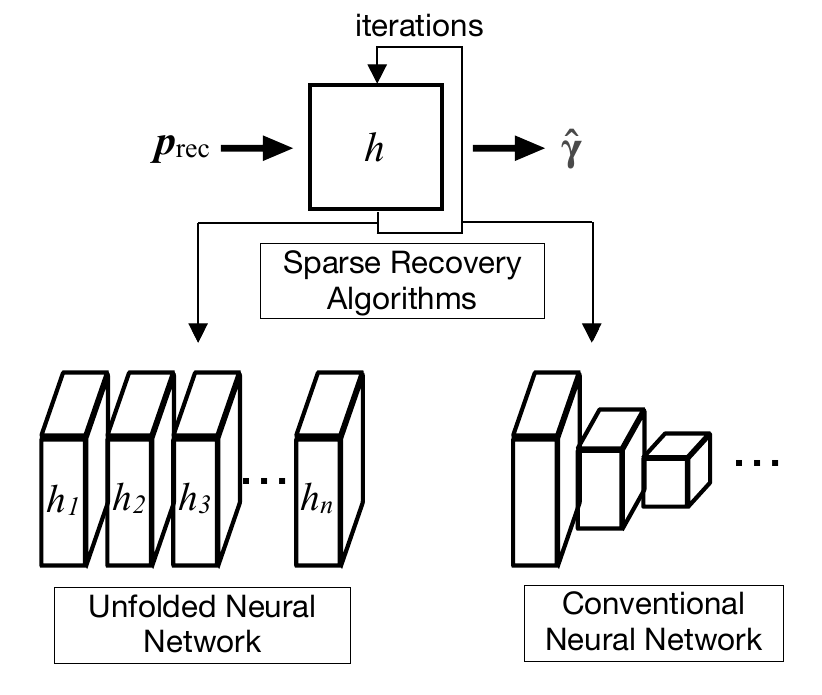}
\caption{\centering Inversion through unfolding of classical algorithms (left) or direct replacement via neural networks (right)}
\end{figure}
 
{Artifact removal by means of \glspl{gan} is studied in \citep{8417964}.} \glspl{gan} consist of two different network structures: one generates data and the other discriminates whether the data belongs to the labeled data or was transferred from the generator network. This process continues until the discriminator network fails to differentiate between the two groups. In \citep{8417964}, the input data fed to the generator network is obtained as ${\boldsymbol{\bar{\gamma}}} = \bm{G}^\dag\bm{p}_{\mathrm{rec}}$ where $\dag$ represents the pseudo-inverse. A \gls{drl} \citep{he2016deep} architecture is employed as a generator network, while the discriminator uses a \gls{cnn} architecture to classify between two groups of images. Additionally, between the generator and discriminator networks, another block is located which tunes the output of the generator network in order to generate a feasible set of images instead of overemphasizing the noise values of the pixels. 

\input{shrink_ultrasound/3.1.2_model_based_DL}

\input{shrink_ultrasound/3.2_tomography}

\input{shrink_ultrasound/3.3_us_classification}

\subsection{Dictionary Learning} \label{sec:ut_dictlearning}

In {ultrasound imaging inverse problems}, selecting an appropriate dictionary for the input data is {crucial} in finding low dimensional representations of data, as well as in inverse problems such as imaging and denoising. Dictionaries can be designed by reducing the mutual coherence of the dictionary matrix, {through} standard optimization, or via learning techniques.

{The $k$-SVD \citep{1710377}, a prominent method for dictionary optimization, is based on \gls{svd}}. It employs an alternating minimization approach, updating one dictionary matrix {column} and one data matrix {row} sequentially, while keeping {other} variables fixed. {Notably}, the $k$-SVD employs an approximate error matrix for updating the column of the dictionary matrix, rather than the actual one. In the study \citep{Song2022}, {this limitation is addressed in the context of sparsifying ultrasound elastography images by using the first principal component from the actual error matrix's \gls{pca}, in contrast to the $k$-SVD's use of the approximate error matrix's \gls{svd}.}  


Wavelet-based representation images is common in \gls{uct} and \gls{mri}, {but the search for suitable dictionaries is ongoing} since images synthesized from wavelets are not necessarily in the category of natural images. In \citep{tovsicultrasound}, authors {explore} a better dictionary representation for ultrasound tomography and \gls{mri}. They employ an optimization method based on \citep{olshausen1996emergence}, which is an iterative alternating minimization method based on the loss function of the basis pursuit denoising problem. {This method alternates between fixing the dictionary matrix and the image data, resulting in a 2dB improvement in \gls{psnr} level}.


\input{shrink_ultrasound/3.5_SyntheticUS}

%% file: shrink_ultrasound/3.1.2_model_based_DL.tex
\subsubsection{Ultrasonic Imaging with Model-Based Deep Learning} \label{model_based_DL}

In ultrasound signal processing, integrating prior knowledge, such as structure and physics, is crucial \citep{drinkwater2006ultrasonic}. Model-based algorithms with domain knowledge, such as message-passing techniques in ultrasound imaging \citep{kim2016ultrasound}, offer interpretability but depend on specific expertise. Conversely, data-driven methods, particularly \gls{dl}, adapt to diverse scenarios without domain-specific knowledge \citep{liu2019deep, van2019deep}, but require extensive data and lack explainability due to their opaque nature. Recently, \emph{model-based deep learning} has been proposed to combine the advantages of both approaches \citep{shlezinger2022model}.

Model-based deep learning, which merges prior models with \glspl{dnn}, builds on ideas from over a decade ago. An early example, the \gls{lista}, incorporated neural network layers into \gls{ista} for sparse signal recovery \citep{gregor2010learning}. A recent survey by Shlezinger explores this field, outlining concepts, design strategies, and applications. The survey categorizes model-based deep learning into two main types: \emph{model-aided networks}, which integrate complete \glspl{dnn} with standard model-based algorithms, such as \emph{deep unfolding} or \emph{algorithm unrolling} \citep{monga2021algorithm}; and \emph{\gls{dnn}-aided inference}, which replaces parts of the original approach with \glspl{dnn}. As an example of \gls{dnn}-aided inference, \citep{bora2017compressed} leverages generative models to map a lower-dimensional vector to an approximation of the sparse vector $\boldsymbol{\gamma}$, aiding subsequent optimization processes.


The integration of model-based deep learning into beamforming represents a significant development in ultrasound imaging. Sloun advanced this field by integrating deep learning methods into various aspects of ultrasound imaging, including scanning modes and digital receive beamforming techniques \citep{van2021deep}. Concurrently, Luijten explored the combination of traditional ultrasound signal processing with model-based deep learning, transforming beamforming into parameter estimation problems \citep{luijten2023ultrasound}. A model-based \gls{dnn} based on \gls{mv} beamforming was developed, achieving high-quality image reconstruction with reduced computational demands \citep{luijten2019deep, luijten2020adaptive}. The resulting neural network architecture, dubbed \gls{able}, was shown to outperform \gls{das} and compete with \gls{mv} beamforming with regards to \gls{cnr}, \gls{fwhm}, and \gls{mae} at a lower computational cost.

Deep unfolding, recognized for its transparency, reliability, and simplicity, is a prominent method in model-based deep learning, as highlighted in \citep{li2021deep}. Originating from the \gls{admire}, this technique has proven effective in improving image quality and suppressing acoustic clutter, as demonstrated in studies by Khan \citep{ khan2022unfolding}. In the context of \gls{ceus}, the novel model-based \gls{dnn} named \gls{corona} has been introduced \citep{solomon2019deep}. \gls{corona} unfolds an iterative algorithm into a fixed-length architecture, enhancing convergence and image quality.

Moreover, \gls{cs} with model-based deep learning has significantly advanced optimal subsampling pattern exploration. By assuming a finite rate of innovation (FRI) structure in beamformed signals from frequency-domain beamforming \citep{chernyakova2014fourier} or compressed beamforming by \gls{coba} \citep{cohen2018sparse}, an efficient model-based \gls{dnn} utilizing a modified \gls{lista} has been developed. This approach, as shown in \citep{mamistvalov2021deep}, considerably reduces array elements, sampling rates, and computational times while preserving imaging quality. Furthermore, Huijben applied the Gumbel-Softmax reparameterization trick with \gls{lista}, introducing the \gls{dps} approach \citep{huijben2020learning} which turns the combinatorial problem of choosing $k$ out of $n$ samples optimally into one that can be solved with gradient-based methods. In contrast, traditional approaches such as the ones explored in \citep{perez2020subsampling} rely on greedy methods and search for suboptimal solutions in order to remain tractable in large scenarios. 

\begin{figure}[ht!]
	\centering
	\includegraphics[width=1\columnwidth, trim = {0cm 0cm 0cm 0cm}, clip]{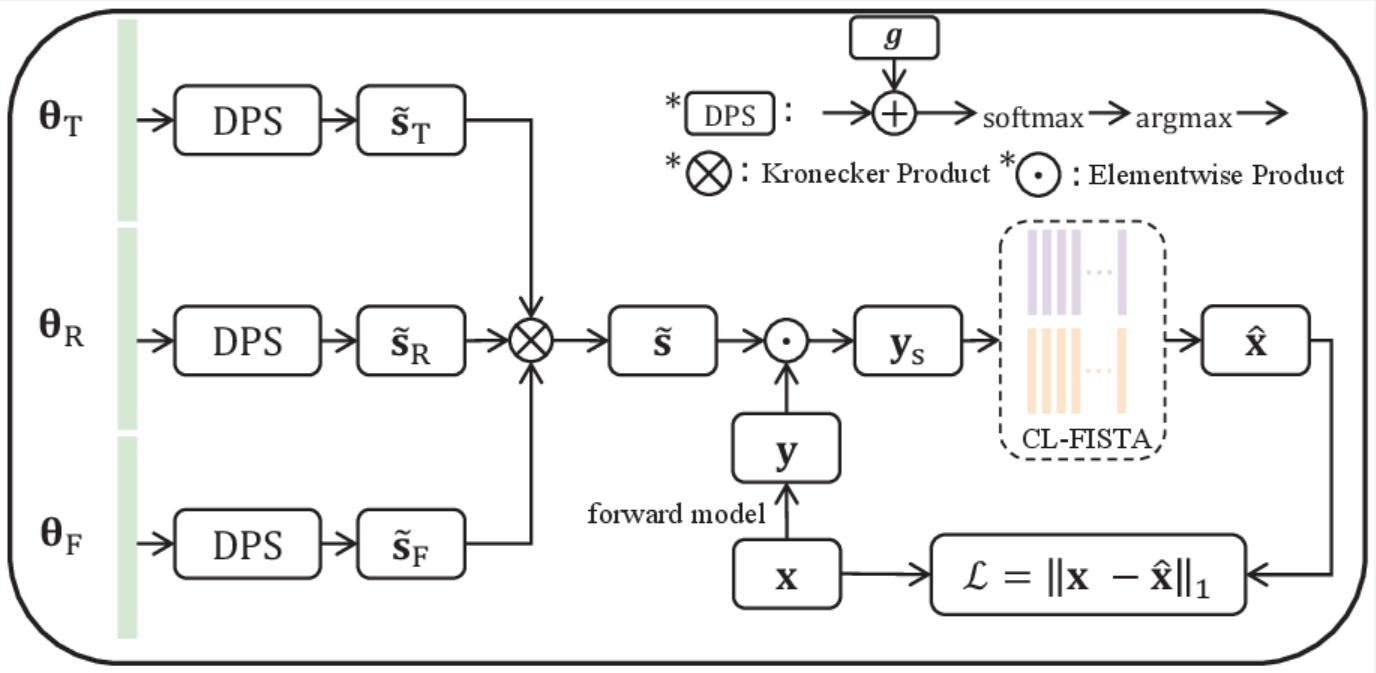}
	\caption{Structure of Joint DPS Algorithm} 
	\label{fig_jdps}
\end{figure}

Within the \gls{dps} framework, an optimal spatial subsampling design for single-channel synthetic aperture imaging has been developed, focusing on maximizing Fisher information \citep{wang2023smsi} or minimizing reconstruction error \citep{wang2023eusipco}. In multichannel imaging, the design of subsampling matrices based on information theory-based and task-based targets was studied within the \gls{dps} framework \citep{wang2023ius}. Additionally, \gls{jdps} has been introduced in \citep{wang2024icassp} for the structured selection of transmitters, receivers, and Fourier coefficients during measurements. The parallel network architecture, illustrated in Figure \ref{fig_jdps}, significantly reduces trainable parameters, enhancing the practicality of the \gls{cs} pattern in hardware applications. 


%% file: shrink_ultrasound/3.2_tomography.tex
\subsection{  {Computed Tomography and Full Waveform Inversion}} \label{section_tomography}

In \gls{uct}, internal cross-sections of an object are derived from ultrasound propagation models and measurements collected from sensors on the object's surface \citep{khairi2019ultrasound} The cross-sections are quantitative in nature, accurately describing the physical parameters of the object under test. Traditional \gls{uct} methods use the first-order Born approximation \citep{gubernatis1977born} and linear ray models, yielding algorithms akin to X-ray imaging. For instance, \gls{art}, typical in X-ray, is applied for slowness maps in masonry pillars, aiding in computing Young's modulus \citep{zielinska2018non}. Similarly, \citep{haach2016qualitative} employs \gls{sirt} for slowness map generation. Nonetheless, these approaches overlook critical phenomena like refraction and multiple scattering, thereby constraining resolution and quantitative interpretability \citep{duric2007detection}. 

The ultrasound community has recently adopted techniques from geophysics and seismology, notably \gls{fwi}. This method emphasizes the accurate solution of \cref{eq_wave,eq_helm} to simulate realistic wave propagation \citep{virieux2017introduction}. It is typically framed as the nonlinear least squares problem

\begin{align} \label{eq_fwi}
    \argmin_{\bm{\gamma}} \left\Vert G(\bm{\gamma}) - \bm{p}_{\mathrm{rec}} \right\Vert_2^2,
\end{align}
where $G$ represents a nonlinear function mapping parameters $\bm{\gamma}$ to simulated data, typically via a differential equation solver. Alternatively, the problem can be reformulated in the frequency domain using $\tilde{\bm{p}}_{\mathrm{rec}}$.

\gls{fwi} has shown promise in medical ultrasound, particularly in high-resolution brain imaging \citep{guasch2020full}. Yet, its implementation is demanding, requiring significant computational resources for precise ultrasound simulation. Moreover, the ill-posed nature of the problem in \cref{eq_fwi} necessitates effective regularization. 

Neural networks are increasingly utilized in tomography with and without \gls{fwi}, either as a component substitute or a complete replacement for traditional methods. For instance, in \citep{klosowski2020maintenance}, neural networks aid in detecting inclusions within reactor tanks. Here, $\bm{\gamma}$ denotes the presence of inclusions. A preprocessing step transforms the measurement data $\bm{p}_{\mathrm{rec}}$ into a time-of-arrival vector $\bm{\tau}$ which is then mapped by a classification \gls{cnn} $\Psi$ to an estimated $\hat{\bm{\gamma}}$. Similarly, \citep{lahivaara2018deep} employs a neural network for inversion in material parameter estimation. They use a modified Galerkin method for simulation and a \gls{cnn} for inversion, focusing on a task-based approach with a cost function that minimizes the difference between predicted and reference material parameters $\bm{\gamma}$. These studies, along with others reviewed in \citep{cantero2022deep}, highlight the growing role of \glspl{cnn} in ultrasonic computed tomography (\gls{uct}) and non-destructive testing (\gls{ndt}). 

\gls{uct} is more common in medical imaging than in \gls{ndt}, yet it remains nascent in medical applications. Nevertheless, progress in integrating \gls{uct} with artificial intelligence is more pronounced in medical imaging. A prime example is \citep{long2023deep}, where authors reconstruct a speed of sound map $\bm{\gamma}$ using a small number of sensors. Differing from previous studies, they apply a network in the image domain: a low-resolution image $\bm{\bar \gamma}$ obtained using a bent ray model is refined using a U-Net \citep{ronneberger2015u}, trained to match high-resolution maps obtained with extensive sensor data. The training focuses on optimizing the reconstruction fidelity $\Vert \Psi(\bm{\bar\gamma}) - \bm{\gamma} \Vert_2^2$. Similar approaches are seen in \gls{ndt}, such as in \citep{feigin2019deep}, where a \gls{cnn} computes $\bm{\gamma}$ from limited plane wave data and forward modeling is done with Matlab's k-wave toolboox. The rising use of \glspl{cnn} in medical imaging underscores their significance in advancing \gls{uct} applications. 

A separate approach to modeling and inversion is provided by \glspl{pinn} \citep{raissi2019physics, moseley2020solving, rasht2022physics}. \glspl{pinn} break away from the previous patterns of using classical methods of forward modeling. Instead, the focus is shifted onto the representation of solutions to the governing differential equations. To this end, a neural network $u_{\bm{\theta}}: (\omega, \bm{x}) \mapsto {p}(\omega, \bm{x})$ with internal parameters $\bm{\theta}$ is employed, noting that a similar formulation is possible in the frequency domain. 

The power of this approach comes from using a simple neural network, usually a shallow \gls{dn} that does not explicitly depend on $\gamma$, endowed with desirable properties through the judicious choice of a cost function. The cost should incorporate the governing equations and all available initial and boundary conditions. As an example in the frequency domain, consider the Helmholtz equation. Based on \cref{eq_helm}, a term of the form
\begin{align} \label{eq_pinn}
    \left\Vert (\nabla^2 + k^2)(\tilde u_{\bm{\theta}}) + \tilde{s} \right\Vert^2
\end{align}
can be included in the cost function. Significantly, \cref{eq_pinn} is expressed in a continuous format, underscoring that $\tilde u_{\bm{\theta}}$ is a solution to the differential equation everywhere in the computational domain $(\omega, \bm{x})$. In practice, a small set of representative points is chosen to evaluate \cref{eq_pinn} at. This acts as a regularizer for the data fidelity term $\Vert \tilde u_{\bm{\theta}}(\omega, \bm{x}) - \tilde{p}_{\text{rec}}(\omega, \bm{x}) \Vert^2$. Similar terms can be formulated for the initial and boundary conditions, or these can be learned implicitly if enough data is available.

In the case of inverse problems, the parameter $\gamma = k$ can be replaced with a neural network $\Psi_{\bm{\phi}}: (\omega, \bm{x}) \mapsto k(\omega, \bm{x})$ with internal parameters $\bm{\phi}$. After an initial training stage for $\tilde u_{\bm{\theta}}$ with data from a known reference medium, one can then train $\tilde u_{\bm{\theta}}$ and $\Psi_{\bm{\phi}}$ simultaneously from measurement data of the object of interest by considering the problem
\begin{align} \label{eq_pinn_inv}
    \argmin_{\bm{\theta}, \bm{\phi}} \left\Vert (\nabla^2 + \Psi_{\bm{\phi}}^2)(\tilde u_{\bm{\theta}}) + \tilde{s} \right\Vert^2 + \lambda \left\Vert \tilde u_{\bm{\theta}} - \tilde{p} \right\Vert^2 
\end{align}
with possible additional terms accounting for boundary and initial conditions. An example of this approach can be found in \citep{shukla2020physics}, where \glspl{pinn} are employed in the estimation of speed of sound maps to locate cracks with high accuracy based on time domain data.

The tomography and \gls{fwi} approaches discussed in this section can roughly be categorized as shown in \cref{tab:tomo}.

\begin{table}[H]
\caption{Classification of \gls{ml}-aided \gls{uct} and \gls{fwi} approaches}
\label{tab:tomo}
\begin{center}
\begin{tabular}{ m{1.8cm} m{2.6cm} m{2.6cm} } 
 \hline
    Approach & Forward method & Inverse method\\
 \hline
 \hline
    Direct & classical & CNN   \\
    Two-step & classical & preprocessing + CNN \\
    PINN & DN & DN \\
 \hline

\end{tabular}
\end{center}
\end{table}

%% file: shrink_ultrasound/3.3_us_classification.tex
\subsection{Ultrasound Data Classification} \label{sec:ut_classification}

In the \gls{ndt} 4.0 era, the fusion of \gls{dl} and \gls{ml} with ultrasound defect classification has markedly enhanced the performance and efficiency of industrial applications, resulting in algorithms that are smarter, more robust and accurate, and more cost-effective. As with many data-driven methods, the success of classification methods in \gls{ut} hinges on the availability and quality of training data \citep{SUN2023106854}; the topic of data augmentation and synthesis is addressed in \cref{sec:ut_datasynth}. This section aims to provide an overview of recent advancements in ultrasound classification within \gls{nde}, highlighting works that integrate feature extraction with classification techniques. Feature extraction is critical in ultrasound, where sound data can be represented in various ways. This step quantifies key data characteristics, aiding classification algorithms in accurately distinguishing between classes. The effectiveness of combining signal processing methods, such as time-frequency-scale signal representations and dimensionality reduction, with \gls{dl} networks and \gls{ml} classifiers, is demonstrated in the examples provided.

The \gls{stft} and wavelet transform, providing simultaneous time and frequency domain information of signals, are integral in ultrasound classification for feature extraction \citep{57199}. Despite \gls{stft} being a traditional technique, its recent applications in \gls{nde}, specifically in structural health monitoring of railway tracks using \glspl{cnn}, are notable \citep{app13010384}. Regarding the identification of rail defects, the wavelet transform has also proved to be a valuable tool for feature extraction. In \citep{JIANG2019455}, wavelet coefficients, energy, and local entropy are utilized for extraction of the representative features of the data alongside \gls{kpca} for the dimensionality reduction algorithm. The authors of \citep{JIANG2019455} propose utilizing \gls{svm} in the classification task. In \citep{LV2023102752}, three  \gls{ml} classification techniques, in particular adaptive boosting \citep{adaboost}, \gls{xgboost} \citep{xgboost} and \gls{svm} \citep{svmc}, are compared by accompanying wavelet energy features for classifying the defects with laser ultrasound signals.

Addressing another crucial \gls{nde} subsector, in \citep{9044747}, \gls{dwt} is applied in a pipeline weld inspection using \gls{emat}. The high-level features from the time-frequency representation of A-Scan signals are used to classify the welding defects in a gas pipeline. The wavelet features are then given into a \gls{cnn}-\gls{svm} classifier. Different solutions to the problem of classifying ultrasound signals in welding applications are proposed in a similar manner by employing the wavelet coefficients-based extracted features in \citep{Chen2023} with \gls{abc}-\gls{svm} and in \citep{pawarpranav} with \glspl{cnn}. Another efficient implementation combining the extracted wavelet coefficients and \glspl{cnn} is presented in \citep{electronics10151772} and \citep{arbaoui2021} for identifying and monitoring the cracks in concrete, respectively.

The impact of the wavelet decomposition on classification accuracy is demonstrated in \citep{Sudheera2020} by comparing two methods: one using \gls{lstm} \citep{hochreiter1997long} directly on the ultrasonic signals, and the other one applying \gls{lstm} after the wavelet feature extraction carried out. Yet another strength of the wavelet transform is extracting definitive features so that they can be used with unsupervised machine learning algorithms such as Gaussian Mixture Modeling, Mean Shift Clustering, and K-means Clustering \citep{8926078}. Furthermore, apart from taking the wavelet coefficient-based values directly as features, the usage of wavelets in reliable feature extraction can be utilized as a preprocessing step as well. In \citep{OLIVEIRA2020106166}, the wavelet transform is carried out to denoise data in the context of detecting the possible damages in wind turbine blades.

Wavelet-based feature extraction is diversified through methods like \gls{ewt}, which adaptively decomposes signals into a finite number of components without requiring prior time-frequency information \citep{6522142}. The authors illustrate the combination of \gls{ewt} with \glspl{cnn} for classifying phased array ultrasonic test signals. \gls{ewt} draws inspiration from \gls{emd}, a method for empirically deconstructing data into intrinsic oscillatory modes, capturing local time-scale signal information \citep{huang1998empirical}. \gls{emd} was applied alongside \gls{dwt} and other features, like entropy and crest factor, for acoustic emission signal-based wood damage and fracture classification, using \gls{lda} for classification \citep{zhang2021classification}. Additionally,\gls{vmd} distinguishes itself by decomposing ultrasound signals through a non-recursive, constrained variational problem-solving approach \citep{vmd}. \gls{vmd} has been applied in ultrasonic measurement classification, such as in wood hole-damage \citep{wood} and for signal denoising prior to feature extraction \citep{9247428}.

Ultrasound signals, when represented sparsely using spatial sparsity and dictionary matrices \citep{schiffner2019random}, show promise in classification accuracy. The authors demonstrate this by replacing traditional iterative sparsity-promoting algorithms with a \gls{cnn} for classifying low-quality, sparse reconstruction data \citep{lohit2016direct}. This transformation involves a simple projection from the measurement to the image domain. The \glspl{cnn}, configured in various architectures like LeNet-5 \citep{726791} and CaffeNet \citep{jia2014caffe}, outperforms the Smashed Filters algorithm \citep{davenport2007smashed}, which uses \gls{mle} for manifold classification, in terms of accuracy.

Shifting from signal processing to purely data-driven methods, a study \citep{8863361} applies a \gls{fcnn} and a \gls{gru} for feature extraction in ultrasound signals, followed by a softmax layer for classification. The authors of \citep{HA2022106637} employ an autoencoder to discern defect-free signal characteristics and identify anomalies upon defect presence. In contexts where ultrasound data is sparsely represented, the authors demonstrate that classifying compressed or subsampled data using soft-margin \glspl{svm}, yields results as accurate as those obtained using the original fully sampled data \citep{Calderbank2009CompressedL}. Notably, this approach obviates the need for data reconstruction, significantly reducing computational complexity, particularly when the original data size substantially exceeds the subsampled one.

Last but not least, a comprehensive analysis of the recent enhancements in \gls{nde} with ultrasound data using numerous \gls{dl} techniques utilized as feature extraction methods, such as autoencoders, \glspl{gru}, \glspl{rnn}, \glspl{cnn} is discussed in \citep{CANTEROCHINCHILLA2022102703}.

%% file: shrink_ultrasound/3.5_SyntheticUS.tex
\subsection{Synthetic ultrasound data} \label{sec:ut_datasynth} 


The generation of synthetic data increasingly attracts interest in \gls{ndt} applications such as \gls{ut}. On one hand, the validation of inspection procedures through sizing and classification of defects as well as \gls{pod} requires large amounts of measurement data representing various types of inspection modalities and defects \citep{virkkunen2023virtual}. On the other, the training of \gls{ml} models requires volumes of labeled data pertaining to a particular application which in practice are difficult to obtain. This section discusses the augmentation and synthesis of measurement data using \gls{ai} techniques.

Pyle explored uncertainty quantification in \gls{dl} for ultrasonic crack characterization, employing a \gls{cnn} to assess surface-breaking defects in \gls{pwi} images \citep{PHA22}. Two uncertainty quantification methods were applied: deep ensembles and Monte Carlo dropout. The network, trained on 14,343 simulated \gls{pwi} images of surface-breaking defects generated via a hybrid \gls{fe}/ray-based model, was further calibrated using a mix of simulated and experimental images. Their findings indicated that while Monte Carlo dropout showed limited effectiveness in uncertainty quantification, deep ensembles demonstrated superior calibration and anomaly detection capabilities. Enhancements to deep ensembles, including spectral normalization and residual connections, further improved calibration and the detection of out-of-distribution samples. Additionally, Gantala explored crack detection using six synthetic \gls{tfm} imaging datasets, created through a combination of \gls{fem} simulation, \gls{gan}-based methods, and realistic noise extracted through the sliding kernel approach, to enhance an \gls{adr} in butt-welds \citep{GB21}. Utilizing a fine-tuned pre-trained YOLOv4 network \citep{yolov4}, they achieved a \SI{100}{\%} \gls{pod} in defect detection and less than \SI{10}{\%} false-calls on the hybrid dataset without noise, and \SI{90}{\%} \gls{pod} with less than \SI{17}{\%} false-calls on the noisy dataset.


Posilovic applied \glspl{gan} to create ultrasonic images that mimic real ones, using a dataset of scanned steel blocks with artificially induced defects \citep{PMS22}. This dataset {consists of} 3825 images, {comprising} a total of 6238 annotations and 4283 defect patches after manual filtering. They explored three synthetic image generation methods: two deep learning-based \gls{gan} approaches and a traditional copy/paste technique. The first \gls{gan} method, detectionGAN, is a U-Net generator integrated with two PatchGAN discriminators \citep{patchgan} and a pretrained YOLOv3 \citep{yolov3} object detector. The second method, SPADE GAN \citep{spadegan}, modifies the pix2pixHD \citep{pix2pixhd} architecture by incorporating a pretrained YOLOv3 discriminator. Their methods were assessed through tests conducted by highly trained human experts.

Furthermore, Singh \citep{STC22} and Ni \citep{NG20} presented novel applications of generative AI in ultrasound imaging and elasticity modulus identification, respectively. Singh developed a \gls{dnn} framework for real-time super-resolution mapping in ultrasonic non-destructive evaluation, using full aperture and pitch-catch transducer configurations to reconstruct crystallographic orientation maps. They employed a \gls{gan} with a U-Net generator and PatchGAN discriminator, achieving a fourfold increase in resolution and up to $\SI{50}{\%}$ improvement in structural similarity. Ni focused on solving the inverse problem in elasticity, utilizing a conditional generative adversarial net to create shear modulus distributions from strain images. Their approach involved a U-Net architecture to implement a conditional GAN generator \citep{cgan}, and a {PatchGAN} classifier for discrimination.

Alternatively, since ultrasound propagation obeys physical laws, solving the wave equation \eqref{eq_wave} is another approach for \gls{ut} data synthesization. Given a reference speed of sound distribution $c(\bm{x})$ and proper boundary conditions, the solution to the wave equation corresponds to the data that would be measured from the corresponding specimen. Conventionally, \glspl{pde} including wave equations are solved using classical numerical methods such as finite elements or spectral approaches \citep{GB21, PHA22}. Due to their robustness and sophisticated algorithms, such numerical methods play crucial roles when it comes to data synthesization. Yet, their high computational cost may become prohibitive especially when dealing with a large 3D region. Since the object surface needs to be specified, dealing with complicated geometry is also challenging with such numerical methods. \par

Outside of their applications in inverse problems such as \gls{fwi} and \gls{uct}, \glspl{pinn} have also shown significant promise purely in forward modeling, i.e. in enhancing numerical methods for solving the wave equation \citep{raissi2019physics}. Moseley demonstrated that a simple 10-layer shallow \gls{mlp} can effectively simulate 2D wave propagation in inhomogeneous media \citep{moseley2020solving}. Song tackled the challenge of balancing physics and boundary loss in \glspl{pinn} by integrating perfectly matched layers into the Helmholtz equation, achieving impressive results in inferring 2D wave fields with anisotropic wave speeds in inhomogenous media \citep{Song2021pinnhelmholtz}. Offen explored a variational approach to \glspl{pinn}, successfully inferring 1D wave propagation in a homogeneous medium with reflective boundaries \citep{Offen2023pinnvariational}. This approach reformulates a second-order \gls{pde} as a first-order differential equation, providing an effective alternative for time-domain wave equation solutions. For a comprehensive overview of \glspl{pinn}, the review by Karniadakis is recommended \citep{Karniadakis21pinnreview}.

%% file: sections/thermography.tex
\section{Thermography} \label{section_thermography}
Infrared thermography has established itself as a valuable technique in non-destructive evaluation due to its ability to provide prompt, swift, and cost-effective information. We will explore both state-of-the-art conventional methodologies and their AI-based counterparts in this field. Leveraging \gls{ai}, thermography effectively eliminates blurring and reconstructs initial temperature profiles. A pivotal focus is defect detection using \gls{dl} architectures, discussed comprehensively in this section.

\subsection{Pulsed Phase Thermography} \label{sec:ppt}
\gls{pt} \citep{almond1994defect} is one of the popular stimulus methods of \gls{tndt}. 
The specimen undergoes a pulse of external thermal stimulus, and the resulting temperature decay is captured by an \gls{ir} camera. This pulse of energy can be generated using high-power amplifiers, flash lamps, laser beams and water jets. The setup can be deployed in two modes: reflection and transmission.
In reflection mode, both the source and the detector are positioned on the same side of the sample. In contrast, in transmission mode, the source and detector are located on opposite sides of the sample to be inspected. The effectiveness of this method varies based on the pulse duration, ranging from a few milliseconds for materials with high thermal conductivity, such as metals, to a few seconds for materials with low thermal conductivity, like wood. This allows for the inspection of different materials using the \gls{tndt} approach.

In a qualitative assessment, when a pulse is applied to the specimen, its temperature rises during the application of the pulse. Subsequently, the temperature decreases as a thermal front propagates beneath the surface, following the principles outlined in the Fourier diffusion equation, which is given by

\begin{equation} \label{fourier_diff}
    \frac{\partial T}{\partial t} = \alpha \nabla^2 T.
\end{equation}

Here, $\frac{\partial T}{\partial t}$ is the is the rate of change of temperature with respect to time, $\alpha$ is the thermal diffusivity of the material and $\nabla^2T$ represents the Laplacian of the temperature field, indicating how temperature varies spatially.
The presence of subsurface defects modifies the diffusion rate, resulting in temperature variations above these defects when compared to the surrounding undamaged areas. Consequently, thermal differences can be detected in thermal images above defect areas. Moreover, the diffusion rate is time-dependent, leading to delayed and less pronounced manifestations of deeper defects. The expressions
\begin{equation} \label{pt_equ}
    \begin{split}
    t \propto \frac{z^2}{\alpha}  \\
    c \propto \frac{1}{z^3}
    \end{split}
\end{equation}
establish the relationship among the observation time $t$, the thermal contrast $c$, the defect depth $z$, and the thermal diffusivity $\alpha$. 

In \gls{ppt} the sample is stimulated as in \gls{pt} technique whereas the analysis is similar to \gls{mt} technique \citep{maldague1996pulse}. In this scenario, a brief rectangular heat pulse is applied to the sample, and its reaction is observed through an infrared camera. This pulse has a wide range of frequencies, which can be separated using the Fourier transform. The data analysis involves choosing a specific frequency, akin to synchronizing the output signal with the modulated input signal, similar to the process in \gls{mt}. The Fourier transform of every pixel (i, j) in the imaging array is then computed along time domain axis. 

The energy of the stimulating pulse is primarily concentrated in the lower frequency range of the spectrum. According to \cref{pt_equ}, lower frequencies can penetrate deeper, while higher frequencies are confined closer to the surface. Hence, it is appropriate to select lower frequencies for analysis. Additionally, it has been observed that the phase image remains unaffected by the surface properties of the material and the non-uniformity of the source, unlike the amplitude image. However, in \gls{ppt} analysis, time information is lost while retrieving frequency data. Combining time information with frequency data can be immensely valuable for depth estimation analysis. Advanced techniques such as wavelet analysis, neural networks, synthetic data, and statistical methods can be employed for such analyses.

In the current literature, investigations have been conducted to detect and evaluate defects by modifying the \gls{ppt} algorithm. In \citep{netzelmann2020modified} two modified algorithms were introduced. The first algorithm integrates a frequency-dependent Gaussian window function, while the second algorithm utilizes a rectangular window function. 
The algorithms underwent testing using synthetic signals that modeled a subsurface cylindrical defect in a plate, and experimental testing was conducted through \gls{pt} on steel and a polymer sample. The outcomes revealed a notably improved contrast-to-noise ratio for defects, particularly at higher analysis frequencies.

In the general context, \gls{ppt} necessitates the assessment of multiple-phase images at various frequencies for a comprehensive sample evaluation. However, the selection and subsequent evaluation of these images are encumbered by certain challenges and are not always straightforward. In efforts to ameliorate these limitations, \citep{poelman2020adaptive} introduced an \gls{asbi} processing technique. This approach surpasses the capabilities of \gls{ppt}, yielding an enhanced signal-to-noise ratio even for defects that are barely visible. Furthermore, it produces a single index map of defects for a given sequence, simplifying the interpretation process.

\subsection{Principal Component Thermography} \label{sec:pct}

\gls{pca} is a widely used multivariate statistical technique designed to reduce the complexity of data. Through linear transformations, \gls{pca} identifies the key features within high-dimensional data and maps them onto a lower-dimensional space. This approach offers several advantages, including straightforward implementation, feature extraction, and noise reduction.

In the realm of \gls{irt}, \gls{pct} \citep{RAJIC2002521, milovanovic2020principal, omar2010combined} employs a method involving the singular value decomposition of the covariance matrix derived from thermograms. This process effectively reduces the dimensions of thermographic data and captures significant data changes. Simultaneously, by projecting the original data onto the directions of maximum variability, \gls{pct} efficiently diminishes noise present in the thermographic data.

\begin{figure}[ht]
\centering
\includegraphics[height=4.2cm, width=8.2cm]{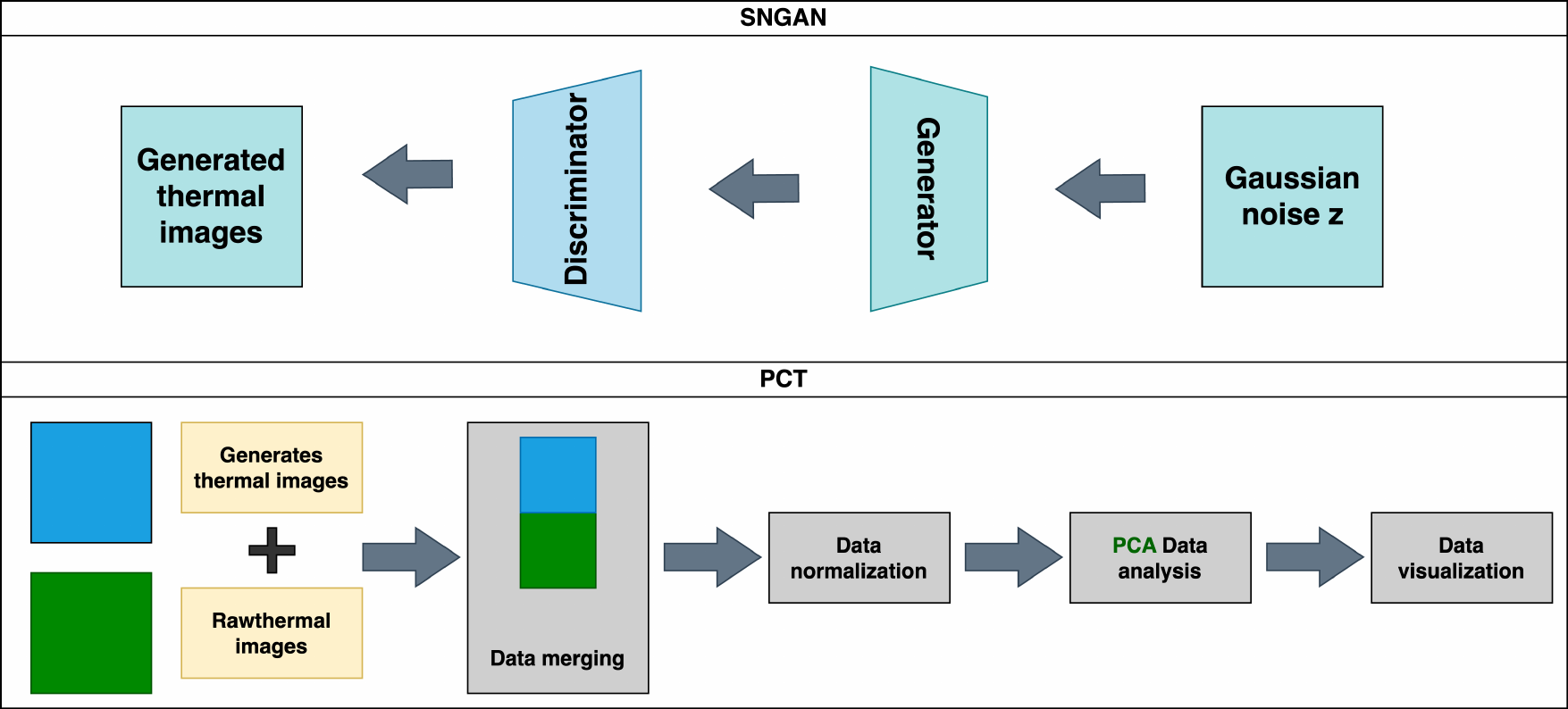}
\caption{\centering Framework of the \gls{gpct} model.}
\label{fig:Thermo-PCT}
\end{figure}

Replying solely on \gls{pct} may not yield satisfactory results. Consequently, numerous defect inspection methods based on deep learning have been proposed in recent times. One such advancement is the \gls{sngan} algorithm \citep{liu2023generative}. The complexity of the generator (G) and discriminator (D) in traditional \glspl{gan} often leads to challenges like vanishing or exploding gradients during training. The \gls{sngan} tackles these issues by incorporating spectral normalization, a technique that normalizes the weight matrix in the discriminator \citep{zhong2023fine}. This approach confines the spectral radius of the weight matrix within a fixed range, effectively preventing issues like gradient disappearance or explosion. By implementing this normalization, the training stability of \glspl{gan} is significantly enhanced, leading to the generation of higher-quality images.
The architecture of the \gls{gpct} model is depicted in \cref{fig:Thermo-PCT}.
In this proposed approach, the traditional \gls{pct} method involves combining the generated images with the original thermal images, followed by \gls{pca} data analysis. 

This integration enhances the overall defect detection process and ensures more accurate and reliable results.

\subsection{Thermographic Signal Reconstruction} \label{sec:therm_reco}
Originally developed for pulse thermography, the \gls{tsr} processing technique \citep{shepard2003advances} consists in the fitting of the experimental log-log plot thermogram by a logarithmic polynomial of degree $n$. Using $\Delta T = \Delta T(t)$ to represent the temperature increase as a function of time (thermogram) for a single pixel $(i, j)$, the temperature increase is modelled as 
\begin{equation}
    \log_{10}(\Delta T) = a_0 + a_1\log_{10}(t) + a_2\log_{10}(t)^2 + \ldots + a_n \log_{10}(t)^n.
\end{equation}
This fitting method replaces the entire sequence of temperature rise images $\Delta \bm{T} (t)$ with a series of $(n+1)$ images $\bm{A}_m$, $0 \leq m \leq n$, representing the polynomial coefficients, so that $\bm{A}_m(i,j)$ is the $m$th coefficient at pixel $(i,j)$. 
These coefficients facilitate the reconstruction of an entire thermographic sequence. The computation of the initial and subsequent logarithmic derivatives of the thermograms is directly conducted on the polynomial, leading to minimal temporal noise augmentation.

The procedures for fitting and deriving thermograms depend on the specific temporal domain being analyzed. It's essential to define a precise time window to isolate the segment of thermograms affected solely by the particular physical phenomena under examination. The traditional use of \gls{tsr} involves selecting optimal derivative images corresponding to distinct depth ranges. Empirically, from \citep{roche2014images} it was determined that a degree of $n=11$ adequately represents delamination-like defects.

In a few recent studies, Schager et al. \citep{schager2020extension} investigated an expanded \gls{tsr} method aimed at automatically categorizing defects found within the undamaged domain.
The approach creates defect maps by utilizing the signal attributes of flaws. The results of the research suggest that the defect map offers an estimate of the defect's depth and is computationally efficient.

Feng et al. \citep{feng2018automatic} introduced a hybrid approach that combines \gls{tsr} with the \gls{asrg} algorithm for thermal image signal processing. The study emphasizes that the selection of defective regions is sensitive to high image resolution, which can enhance detectability. The study showcased enhanced detection resolution and proved valuable for automatically selecting regions suspected of harboring defects.
Ratsakou et al. \citep{ratsakou2020shape} proposed a novel approach for defect characterization, blending \gls{tsr} with the Canny shape-reconstruction algorithms.
The method begins by fitting the raw thermographic images to a low-degree polynomial in the log-log representation of the time axis. Subsequently, the Canny algorithm is employed to reconstruct the original signals. By harnessing the capabilities of the \gls{tsr} algorithm for signal denoising and compression, the method can improve the efficiency of the reconstruction process.

\subsection{Thermographic Image Reconstruction}


The techniques mentioned in the preceding sections do not exploit spatial information and physical knowledge of the problem simultaneously, resulting in limited spatial resolution. In the case of \gls{tsr}, each pixel is traditionally processed separately. In \gls{pct}, spatial information is considered only in the context of forming low dimensional vector spaces. An alternative approach is to perform thermographic image reconstruction by directly addressing the heat equation in \eqref{fourier_diff} and solving an \gls{ihcp} to find the initial temperature distribution in the object under test. Such an approach is comparable to \gls{fwi} in ultrasound imaging and comes with similar challenges, e.g. considerable computational cost and ill-posedness \citep{groz2019three}. 

A recent advancement in thermographic imaging is the transformation of the heat conduction problem into a wave propagation one by means of virtual waves \citep{burgholzer2017three}, after which the usage of the reconstruction techniques covered in \cref{section_ultrasound} becomes possible. In particular, synthetic aperture techniques address both the limited spatial resolution of \gls{tsr} and \gls{pct}, as well as the ill-posedness of \glspl{ihcp}: the incorporation of a simplified model of wave propagation provides robustness and improves to spatial resolution. The inverse mapping from the measured temperature profile $T(t, \bm{x})$, $\bm{x} \in \real^d$, to the desired initial temperature distribution $T(0, \bm{x})$ is performed through an intermediate virtual wave field $p(t, \bm{x})$. Classical and \gls{ai}-based methods can be employed at any step of the process, resulting in increased flexibility. The authors of \citep{DL_thermo_imaging} exemplify this flexibility by comparing three approaches to thermographic imaging: virtual waves with classical ultrasound imaging methods, an end-to-end U-net $\Psi: T(t, \bm{x}) \mapsto T(0, \bm{x})$, and a two-step approach in which the virtual wave mapping $T(t, \bm{x}) \mapsto p(t, \bm{x})$ is done in preprocessing and later fed to a neural network $\Psi:p(t, \bm{x}) \mapsto T(0, \bm{x})$. The authors show that both \gls{ai}-based approaches outperform the classical one, and that the usage of virtual waves in preprocessing increases reconstruction accuracy in exchange for increased inference cost.

\subsection{Defect Shape Detection}

Neural networks for semantic segmentation prove valuable across diverse tasks, such as autonomous vehicles, medical image diagnostics, and other combined applications\citep{gillsjo2022semantic, zhou2023semantic}. In the case of thermographic imaging, semantic segmentation allows the detection and localization of abnormalities, e.g. tumors in medical imaging \citep{inproceedings} and defects in \gls{nde} \citep{semantic_seg_pedrayes}.


In \citep{Defect_shape_detection}, semantic segmentation is performed on induction \gls{ppt} images of forged steel parts by using a U-net. The network is trained by using binary masks denoting the presence and location of cracks. The authors note that, since the cracks are comparatively small in the \gls{roi}, imbalances in the classes must be addressed through weighting in the loss function. As is common in the \gls{nde} context, augmentation through rotations, contrast modifications, and elastic transformations was necessary.




%% file: sections/optical_camera.tex
\section{Optical Camera} \label{opt._insp.}

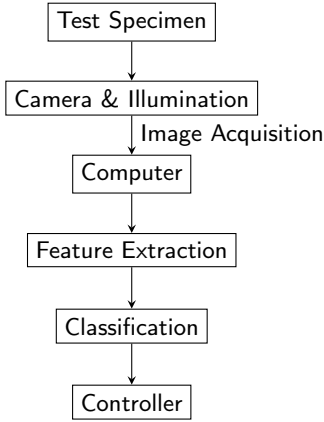
\begin{figure} [h]
  \centering
  \begin{tikzpicture}[>=stealth, node distance=1 cm]

    \node (item) [draw, rectangle] {Test Specimen};
    \node (ki) [draw, rectangle, below of=item] {Camera \& Illumination};
    \node (computer) [draw, rectangle, below of=ki] {Computer};
    \node (feature) [draw, rectangle, below of=computer] {Feature Extraction};
    \node (classification) [draw, rectangle, below of=feature] {Classification};
    \node (controller) [draw, rectangle, below of=classification] {Controller};

    \draw[->] (item) -- (ki);
    \draw[->] (ki) -- (computer) node[midway, right] {Image Acquisition};
    \draw[->] (computer) -- (feature);
    \draw[->] (feature) -- (classification);
    \draw[->] (classification) -- (controller);

  \end{tikzpicture}
  \caption{Setup of a camera system}
  \label{fig:camsys2}
\end{figure}

A significant fraction of \gls{ndt} tasks can be solved through optical inspection methods, i.e., based on optical sensors using image processing methods. Such optical systems can consist of cameras, lasers, scanners, or a combination of the above. However, machine vision with a camera is the most common. \Cref{fig:camsys2} shows the typical workflow of a camera system  \citep{9214824}.

A camera and a suitable (active or passive) illumination are used to take one or multiple images of the specimen, which are subsequently transferred to the computer. 
The image matrix $\bm{P}$ for a grayscale image can be represented as $m \times n$ matrix where each entry $\bm{P}(i, j)$ represents the brightness value of the pixel located at the $i$-th row and $j$-th column of the image. {For color images, each color plane is described like this and written into a three dimensional array.}
Afterwards, feature extraction is applied. Feature extraction in an image can be mathematically represented as a function $f(\bm{P})$ that transforms the original image into a set of features or vectors:
\begin{equation}
    f(\bm{P}) = \{\bm{f}_1, \bm{f}_2, \ldots, \bm{f}_k\}
\label{eq:feature vector}
\end{equation}
Here, $\bm{f}_1, \bm{f}_2, \ldots, \bm{f}_k$ represent the extracted features. These features can be of various types, depending on the type of analysis. {In \gls{nde}, relevant features could be e.g. the edges of a scratch, certain color characteristics indicative of corrosion, etc.} Then the data can be classified. The classification process can be mathematically represented as a function $c(f(\bm{P}))$ that maps the features to a specific class or label:
\begin{equation}
    c(f(\bm{P})) \in \mathcal{C}
\end{equation}
Here, $\mathcal{C}$ is the set of all classes under consideration, and $c$ yields a class or label for the image or region based on the extracted features. {According to the above examples, the question can be clarified whether, for example, the edges found have the characteristics of a scratch, or whether the color features found correspond to those of corrosion.} After the data evaluation, the process chain ends with a controller, which makes a certain decision depending on the class.\\

\begin{figure}[h]
    \centering
    \begin{tikzpicture}
        \node[rectangle split, rectangle split parts=2, draw, align=center] (input) {
            \textbf{Input Layer}
            \nodepart{two} Input Image
        };

        \node[rectangle split, rectangle split parts=4, draw, right=0.4cm of input, align=center] (conv) {
            \textbf{Hidden Layer}
            \nodepart{two} Convolutional
            \nodepart{three} Pooling
            \nodepart{four} Fully Connected
        };

        \node[rectangle split, rectangle split parts=2, draw, right=0.4cm of conv, align=center] (output) {
            \textbf{Output Layer}
            \nodepart{two} Predicted Classes
        };


        \draw[->] (input) -- (conv);

        \draw[->] (conv) -- (output);

    \end{tikzpicture}
    \caption{Convolutional Neural Network (CNN) Architecture}
    \label{fig:cnn_architecture2}
\end{figure}
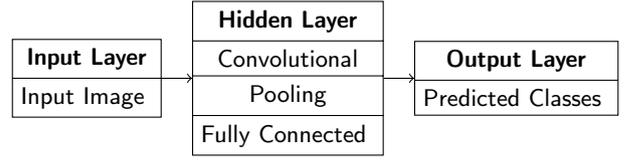

Nowadays, thanks to the advanced development of neural networks, many of these problems can be solved via Deep Learning. A simple scheme is shown in  \cref{fig:cnn_architecture2}.

In the example, the input image goes through many different convolution operations, a common approach in image processing to establish shift invariance of detected objects across the image. During this process, corresponding feature maps are created. The feature maps are compressed by the pooling layer. The fully connected layer connects the different maps so that the desired classification is obtained in the end.

\subsection{Image Preprocessing}

Image preprocessing means the intention to achieve image enhancement in conjunction with formatting the image if necessary. Such operations could include smoothing the image, adjusting the contrast, resizing the image, formatting in a different color range, eliminating the background and more. The various operations depend very much on the actual application. This section briefly introduces the most common operations in general. For deep learning, image preprocessing is essential for technical, qualitative and performance reasons.

Smoothing filters \citep{demant2011industrielle} are used to suppress noise and individual disturbing pixels. A distinction is made between linear and non-linear filters.
Linear smoothing filters \citep{demant2011industrielle} can be described as
\begin{equation}
    \bm{G}(x, y) = \frac{1}{s} \sum_{r=0}^{R-1} \sum_{c=0}^{C-1} f_{cr}\bm{P}(x - d_x + c, y - d_y + r),
\end{equation}
where $\bm{G}(x, y)$ is the filtered image at position $(x, y)$, $\bm{P}(x - d_x + c, y - d_y + r)$ is the pixel value of the input image at position $(x - d_x + c, y - d_y + r)$, which is weighted by the filter coefficient $f_{cr}$. The quantity $s$ is a normalization constant and $d_x$ and $d_y$ are parameters which control the size of the filter. $R$ is the number of rows and $C$ the number of columns in the image. The row and column indices are represented by $r$ and $c$, respectively.

Non-linear smoothing filters \citep{demant2011industrielle} are 'Rank-Order Filters'. Here, a certain number of neighboring pixels are sorted into a row depending on their intensity. A specific pixel is then selected. A common Rank-Order Filter is the so-called Median filter \citep{demant2011industrielle}, which always selects the median intensity and thus makes disturbing pixels disappear completely.

Regarding image enhancement, there are many different ways to improve the contrast of images. A general solution for this is the so-called histogram normalization which can be described via
\begin{equation}
    \bm{G}(x, y) = \frac{L-1}{MN} \sum_{i=0}^{L-1} h(i)
\end{equation}
where $\bm{G}(x, y)$ is the enhanched image, $L$ is the number of intensity levels, $M$ is the image width and $N$ the height. Additionally, $h(i)$ is the count of the $i$-th intensity in the original image. {For color images, these procedures are carried out for each individual color plane.}

Moreover, for deep learning, image resizing is very important, as neural networks need the input images in to be in a homogeneous format. By a scaling factor $s$ the dimensions of the resized image can be calculated based on the original image. Then by applying an interpolation technique the old pixel coordinates can be mapped into the new image coordinate system.

As fully connected layers in convolutional neural networks require the same format for all images, also a color transformation may be necessary. In general we can describe it as follows
\begin{equation}
    \bm{G}(x, y) = \bm{T}[\bm{P}(x, y)]
\end{equation}
$\bm{T}$ is the transformation function which applies a transformation to the color values of each pixel. The transformation function can be defined based on the requirements of the wished color format.

\subsection{Feature Extraction and Classification of Optical Images}
\label{sec:feature extraction and classificsation}

The first step in feature extraction is to define a region of interest in which to search for the desired features. This can be achieved, for example, using threshold methods, edge detection techniques or cluster algorithms. Representative features that characterize the image content are then selected. These can be color information, edges, shapes, textures or specific patterns. The extracted features are then described as a vector as in \cref{eq:feature vector}. The vectors are then filtered again to reduce the amount of data or to discard irrelevant features. Finally, the vectors may also need to be normalized to make them comparable. Then the extracted features can be analyzed so that based on the application the images can be classified.

\begin{figure}
  \centering
  \begin{subfigure}{0.2\textwidth}
    \includegraphics[width=\linewidth]{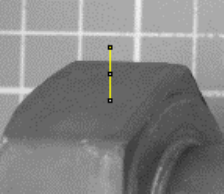}
    \caption{head of screw}
    \label{fig:head of screw}
  \end{subfigure}
  \hfill
  \begin{subfigure}{0.35\textwidth}
    \includegraphics[width=\linewidth]{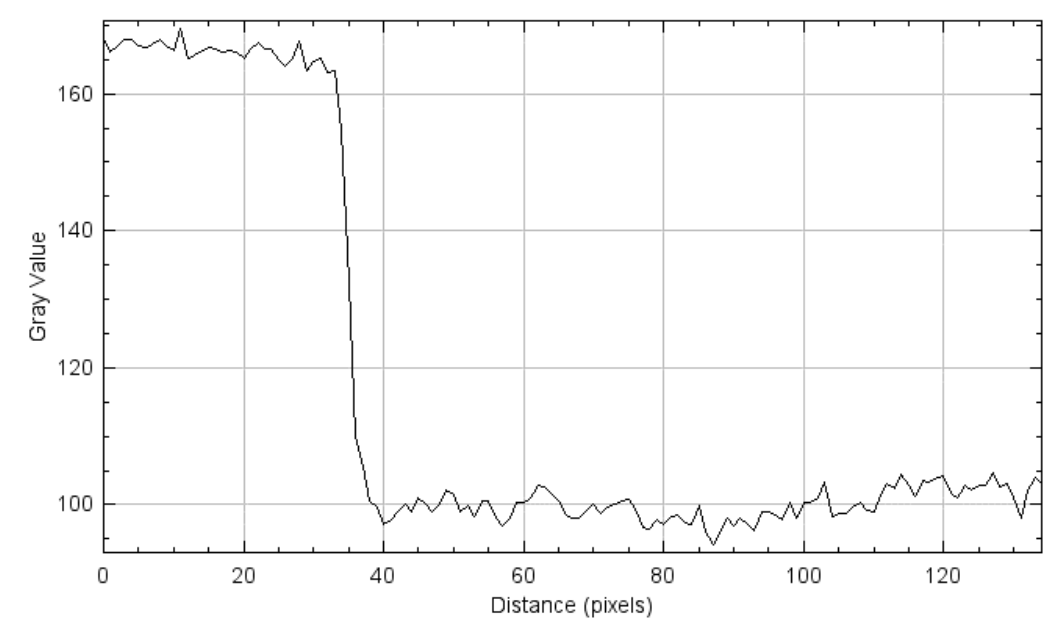}
    \caption{line profile}
    \label{fig:line profile}
  \end{subfigure}
  \caption{Example of feature extraction}
  \label{fig:zwei_bilder}
\end{figure}

\Cref{fig:zwei_bilder} exemplifies feature extraction in optical images by considering a section of the head of a screw. A line is plotted here and the adjacent graphic shows the corresponding line profile. The corresponding drop in intensity from the line function represents the edge of the object and so a corresponding feature can be extracted. If the entire screw is scanned in this way, the contour of the object can be determined and the object can be classified as a screw accordingly. {In the context of Deep Learning, \glspl{cnn} can be trained to register such patterns automatically.} In the following subsection, we review the application of \gls{dl} to optical images in detail.

\subsection{Deep Learning based Feature Extraction and Classification}\label{sec:optic_dl}

Deep learning was revolutionary in image processing. By training neural networks, they can acquire an intuition that is similar to humans' and thus solve highly complex tasks. Image files are usually processed with convolutional neural networks. These networks are predestined for extracting features from images and then classifying them. Tasks such as object recognition, segmentation, finding key points and much more can be easily solved using convolutional neural networks. 

There are many different architectures for this, \glspl{rcnn} \citep{bharati2020deep} among them. A \gls{rpn} is used to find regions that are likely to contain objects. A \gls{cnn} architecture is then used to classify and fine-tune the bounding boxes. Ren et. al \citep{ren2016faster} introduces the Faster \gls{rcnn}. They use as \gls{cnn} architecture the VGG-16 \citep{simonyan2015deep} for the classification. They tested it on the PASCAL VOC 2007 and 2012 \citep{Everingham15} dataset. The best results were achieved when the model was trained with the COCO dataset \citep{lin2015microsoft} and the PASCASL VOC 2007 and 2012 dataset. For the VOC 2007 test data a mAP of 78.8 \% were achieved and for the VOC 2012 test data the mAP amounted to 75.9 \%.

Another architecture is \gls{ssd} \citep{liu2016ssd}. This model uses the VGG-16 architecture as base, but disposes of the fully connected layers. Instead of these, further convolutional layers are added so that several feature maps with different resolutions can be generated. All these maps are taken into account for classification. This not only makes it possible to reliably find and classify the objects, but also to match the object shape when generating the bounding boxes due to the large number of feature maps at different levels. Liu et. al \citep{liu2016ssd} introduced this model. In their paper they differentiate between SSD300 and SSD512, where the number indicates the input size of the images (300x300 vs 512 x 512). Here too, the models were tested with the PASCAL VOC 2007 and 2012 dataset. Again the best results were achieved if the models were trained with the VOC 2007, VOC 2012 and COCO dataset. The SSD300 achieved a mAP of 79.6 \% for the VOC 2007 testset and a mAP of 77.5 \% for the VOC 2012 testset. The SSD512 achieved better results due to the higher resolution of the input images. For the VOC 2007 data the mAP amounts to 81.6 \% and for the VOC 2012 testset it is 80.0 \%.

A widely used method for real-time object recognition is YOLO \citep{jiang2022review}. Here, the image is divided into grids and a prediction is made for each grid. YOLO is based on \glspl{cnn} (see \Cref{fig:cnn_architecture2}). In 2015, Redmon et al. \citep{redmon2016look} introduced YOLO for the first time. They trained the model with the VOC 2007 and 2012 dataset. For the VOC 2007 data it then achieved a mAP of 63.4 \%- and for the VOC 2012 dataset the YOLO scored 57.9 \% mAP. Over the years, new and improved versions of YOLO were published. YOLOv8 \citep{jocher2023yolo} is the latest release. The largest version of the model (YOLOv8x) achieved the best results. It was trained with the COCO 2017 dataset and achieved a score of 53.9 \%. Moreover YOLOv8 also provides models which perform only classification (without bounding boxes), instance segmentation and pose estimation.

In addition to object recognition, image segmentation is also often required. This can be achieved by extending the object recognition models. An example of this is Mask \gls{rcnn}. Mask \gls{rcnn} \citep{he2018mask} \citep{bharati2020deep} is an approach to Instance Segmentation. It uses the Faster \gls{rcnn} \citep{ren2016faster} \citep{bharati2020deep} architecture for object detection and a \gls{fcnn} \citep{7478072} for semantic segmentation. He et al. trained and tested a Mask \gls{rcnn} model with the COCO dataset. The model reached a score of 37.1 AP. In comparasion to that the YOLOv8x-seg model achieved a mAP score of 43.4. This model was trained and tested with the COCO 2017 dataset.

{All these applications are extremely valuable in \gls{nde} tasks. For instance in Katsamenis et. al \citep{katsamenis2020pixellevel}, corrosion detection is carried out. Among other models, Mask \gls{rcnn} is used here. Transfer learning was used to train the Mask \gls{rcnn} model. The Mask \gls{rcnn} model is based on Inception V2 \citep{szegedy2015rethinking} and was pre-trained on the COCO dataset. Then a dataset consisting of 116 images was used to train and test the model. The images were captured using various devices and have different resolutions. Furthermore, the images contained different types of corrosion. The precision achieved was just under 0.7 and the F1-Score achieved was just over 0.7. However, the results were improved by combining them with color segmentation to apply boundary refinement. This shows the importance of classical techniques such as those in \Cref{sec:feature extraction and classificsation} in order to combine AI results with such procedures if necessary.}

%% file: sections/overview_shrink.tex
\section{Common Patterns in Algorithms and \gls{ml} Models} \label{sec:overview}
So far we presented both classical and \gls{ml} based approaches separately for each inspection modality. In the course of investigation, however, we repeatedly encounter some algorithms and/or \gls{ml} architectures regardless of the inspection modalities or even application fields. 
For instance, \glspl{cnn} are ubiquitously employed for computer vision based decision making tasks in many inspection modalities including the ones that are not covered in this paper: defects and/or substance classification in \gls{ut} \citep{app13010384, 9044747} and terahertz inspection \citep{wang2021convolutional}, edge detection in X-ray \citep{xiao2021development} or semantic segmentation in thermography \citep{semantic_seg_pedrayes}, feature extraction and classification in optical images \citep{zaidi2022survey} and X-ray tomography \citep{van2022nondestructive, hena2023deep}. 
Often integrated as a part of feature extraction, \glspl{cnn} can also be used for imaging tasks such as for reconstructing the raw terahertz measurement data \citep{jiang2022machine, helal2022signal} or enhancing the resolution of \gls{ut} \citep{mamistvalov2022deep, perdios2021cnn} and X-ray reconstructions \citep{jiang2022machine, helal2022signal}. 
Such ubiquity can also be found for classical algorithms as well.
An example is correlation based imaging, which is however called differently depending on the applications: \gls{saft} and \gls{tfm} for \gls{ut}, Synthetic Aperture Radar for radar and \gls{das} beamforming for medical ultrasound. \par

All these examples make it evident that there are common patterns in how to process the data depending on the tasks instead of inspection modalities or application fields. 
One can find such patterns, when these tasks deal with the intrinsically same mathematical problems. As a result, these tasks can be formulated mathematically in the similar, if not the same, manner.
For example, reconstructing the inner structure of a test object from \gls{ut} signals can be formulated in a very similar manner as that of reconstructing X-ray \gls{ct} data or image deblurring \citep{Schuler16Deblurr}. This is because all of these tasks aim to recover the quantities which cannot be observed directly from the observations that are affected by these unknown quantities. To be more specific, these unknown quantities can be the location of defects in a test object for \gls{ut} or X-ray reconstruction, and the unknown clean image for deblurring tasks. 
This realization motivates us to make high level connections of the references in terms of their intrinsic natures. For this purpose, we identified four intrinsic mathematical problems we commonly encounter in \gls{ndt} applications, which are \textit{forward problems}, \textit{inverse problems}, \textit{optimal design} and \textit{pattern recognition}. \par

In the following subsections, each problem is described to help identify the intrinsic nature of a specific engineering task. The emphasis is placed on the common patterns one can see when dealing with these tasks. 
In \cref{tab:overview}, some recommended references are provided as a summary of this section. The left most column shows the mathematical problems these references intrinsically solve. The specific \gls{ndt} related tasks are listed in the second column. The third column of the table provides the list of the sections which discuss how these problems are solved for a specific inspection modality. The fourth column shows common methods to solve these tasks, and the fifth column is the list of references for the corresponding methods. 
Furthermore, these tasks are highlighted with four different colors to illustrate how they relate to actual \gls{ndt} inspections. Inspired by the \gls{ndt} automation levels propsed in \citep{cantero2022deep}, we decompose a \gls{ndt} inspection chain into three steps: measurement operation, enhancement of raw measurement data and decision making. In addition to that, we consider data synthesization as an essential part of the future of \gls{ndt} inspections, since there is a pressing need for high quality, well-labeled data for reliable \gls{ai} training. 
Admittedly, this is by no means an exhaustive collection, yet it will give a brief overview of how one can interpret and possibly solve these tasks regardless of applications. 
Note that the conceptual connections we made here is just one possibility, and they can be described and grouped differently. \par

The benefits of \cref{tab:overview} are twofold.
First, acknowledging such conceptual patterns lets us understand the similarities in the tasks and try a method in other modalities or application fields. In the era of \gls{ndt} 4.0, where high level problems need to be solved by machines instead of human operators, we will increasingly encounter the problems for which no state of the art exist. However, once the nature of the tasks is identified, such common patterns can help us determine which measure, algorithm and/or ML architecture is suitable. 
Second, the number of listed methods in \cref{tab:overview} also indicates which tasks are well investigated and which are still in its infancy. This can be useful for determining the direction of future research.

\subsection{Forward Problems} \label{sec_forwardprob}
For the purposes of this work, we refer to general problems of the form $\bm{y} = f(\bm{x})$, which shows a relation between the causal factors $\bm{x}$ and their effects $\bm{y}$ using an operator $f$. For the sake of simplicity, in this work we represent both $\bm{x}$ and $\bm{y}$ as vectors, however they can be arrays as well. \par

Forward problems seek to obtain the effects $\bm{y}$ from the known causal factor $\bm{x}$.
In the context of \gls{ndt}, $f$ usually represents a measurement process, yielding a synthesized version of measurement data $\bm{y}$. 
Especially for \gls{ml}-based approaches, where the lack of quality data has been frequently identified as a major bottleneck regardless of their tasks \citep{cantero2022deep}, solving forward problem is beneficial as it can enlarge the training data set. 
For most cases with \gls{ndt} applications, $\bm{x}$ and $\bm{y}$ are related through the underlying physical phenomenon described by a \gls{pde}, such as Maxwell's equations, wave or heat equations.
Since solving these \glspl{pde} is typically very challenging, the true causal relation is generally approximated. As a result, there are various possibilities to select the operator $f$. 
For accurately computing $\bm{y}$, one can aim to find $f$ such that it approximates the underlying \gls{pde}. Another approach is to simplify the \gls{pde} by making assumptions on the underlying physics, enabling to explicitly compute $f$ to compute $\bm{y}$. 

\subsection{Inverse Problems} \label{sec_inverseprob}
Given the causal relation $ \bm{y} = f(\bm{x})$ we considered in the previous section, inverse problems aim to find the parameter $\bm{x}$ from a set of observations $\bm{y}$ collected through a (measurement) process $f$. 
For example, when $f$ adds noise to $\bm{x}$ so that $\bm{y}$ is a noisy version of $\bm{x}$, the inverse problem is a denoising problem. 
If $\bm{y}$ is an image and one seeks to increase its resolution, $f$ can be interpreted as \emph{subsampling} operation, and recovering $\bm{x}$ becomes super-resolution task. 
In the case of imaging, we consider $\bm x$ to be an \emph{image} in the sense that it describes a parameter that varies in space, and the raw measurement data $\bm y$ is defined in a different domain and/or cannot be interpreted as an image. As a concrete example, eddy current testing deals with the measurement of magnetic field intensity and phase obtained with varying excitation frequencies. In the case of ultrasound, raw data often consists of time domain acoustic pressure signals collected with different measurement channels. For both, the data is processed to obtain an image portraying defect locations. In the case of inverse problems, \glspl{nn} facilitate the computation of $\bm x$ by replacing parts of classical algorithms to invert $f$, or replacing them entirely, i.e. $\bm x = \Psi(\bm y)$, where $\Psi$ is the \gls{nn}. Typically, the quality of the inversion is measured through a cost function $h( \bm{x}, \Psi(\bm{y}))$ which we want to minimize, such as \cref{inverse}.

\subsection{Optimal Design} \label{sec_optdesign}
Optimal design deals with the tuning of experimental parameters which can be freely chosen, and which impact the information content of measurement data with regards to parameters of interest. Recall the inverse problem from \cref{sec_inverseprob}. In reality, the measurement data $\bm y$ depends not only on $\bm x$, but also on the experimental parameters (e.g. the location, number, and type of sensors employed) $\bm \phi$. When solving an inverse problem, the achievable cost $h$ is affected by the experimental design. We can tune the parameters $\bm \phi$ so that the cost $h$ is minimal, i.e. 
$%
\min\limits_{\bm{\phi}} h(f(\bm x, \bm \phi)).
$%

\subsection{Pattern Recognition}
Pattern recognition can be found ubiquitously in many data analysis and decision making tasks. In general, given the input data $\bm x$, we would like to assign it a category or class $y$. There exists a correspondence between $\bm x$ and $y$ given by an unknown mapping $g$ that we wish to find. Whether a category is assigned to $\bm x$ as a whole, by regions, or by sample, provides the nuance among classification, detection, and segmentation, respectively. Oftentimes, the data $\bm x$ is preprocessed through a judiciously chosen function $f$ which highlights informative \emph{features}. This leads to the overall pattern recognition task 
$y = g(f(\bm x))$
which renders itself to a problem of finding an appropriate classifier $g$, preprocessing scheme $f$, or both. \Glspl{nn} can be used to replace the classifier $g$ in combination with classical feature extraction methods such as \gls{pca}, or they can replace both the feature extraction and classification process altogether. 
Since the learning is typically based on the available samples, the quality of the samples determines the performance. 

\begin{table*}
{
\centering
    \scalebox{1.0}{\input{tikz_tables/table_overview.tex}}
    \caption{
        Summary of the methods used for common \gls{ndt} tasks. %
        References are categorized according to the intrinsic mathematical problem they are dealing with as discussed in \cref{sec:overview}. %
        Each color in the table is associated to a \gls{ndt} inspection process: pink for data synthesization, green for data enhancement, orange for measurement operation and blue for decision making. 
        Note that end-to-end \gls{ml} methods often perform feature extraction implicitly. These methods are not shown as feature extraction, instead categorized into their targeted decision making tasks. 
    }
    \label{tab:overview}
}
\end{table*}

%% file: tikz_tables/table_overview.tex
\begin{NiceTabular}{ccccl}[hvlines]
    \CodeBefore
        \rectanglecolor{\ColSynth}{2-2}{4-5} 
        \rectanglecolor{\ColEnhance}{5-2}{15-5} 
        \rectanglecolor{\ColOperation}{16-2}{19-5} 
        \rectanglecolor{\ColEnhance}{20-2}{21-5} 
        \rectanglecolor{\ColDecision}{22-2}{37-5} 
    \Body
        Intrinsic problem & Task & Section & Method & References\\
         \Block{3-1}{Forward problems} 
            & \Block{3-1}{Data\\ synthesizaiton} 
            & \Block{3-1}{\ref{sec:ut_datasynth}, \ref{sec:pct}} 
            & numerical solver & \citep{GB21}, \citep{PHA22}\\
            & & & \gls{pinn} & \citep{raissi2019physics}, \citep{moseley2020solving}, \citep{Song2021pinnhelmholtz}, \citep{Offen2023pinnvariational}\\
            & & & \gls{gan} & \citep{GB21}, \citep{NG20}, \citep{PMS22}, \citep{liu2023generative}\\
        \Block{11-1}{Inverse problems} 
            & \Block{8-1}{Imaging} 
            & \Block{8-1}{
                \ref{sec:eddy_current}, \ref{sec:berkhausen},\\
                \ref{sec:3ma2}, \ref{sec:3max8}, \\
                \ref{sec:ut_imaging_vanillaDL}, \ref{model_based_DL}, \\
                \ref{section_tomography}, \ref{sec:therm_reco}
                }
            & classical & \citep{daubechies2004iterative}, \citep{4959678}, \citep{boyd2011distributed}, \citep{zielinska2018non}, \citep{haach2016qualitative}, \citep{virieux2017introduction}, \citep{guasch2020full}, \citep{balageas2015thermographic}\\ 
            & & & \gls{mlp} & \citep{RTP97}, \citep{GMB99}, \citep{HFL16}, \citep{HRC18}, \citep{ZLL19} \\
            & & & \gls{svr} & \citep{GRK21} \\
            & & & \gls{hlr} & \citep{GRK21} \\
            & & & \gls{xgboost} & \citep{SW23} \\
            & & & \gls{cnn} & \citep{feigin2019deep}\\ 
            & & & Autoencoder & \citep{zhang2020self}, \citep{reconstr_deepl}\\
            & & & \gls{pinn} & \citep{Karniadakis21pinnreview}, \citep{raissi2019physics}, \citep{shukla2020physics}\\
            & \Block{3-1}{Super-resolution} 
            & \Block{3-1}{\ref{sec:ut_imaging_vanillaDL}, \ref{model_based_DL},\\
            \ref{section_tomography}, \ref{sec:ut_datasynth}} 
            & \gls{cnn} & \citep{feigin2019deep}\\
            & & & U-Net & \citep{long2023deep}, \citep{ronneberger2015u}\\
            & & & \gls{gan} & \citep{8417964}, \citep{STC22}, \citep{DL_thermo_imaging}\\
        \Block{4-1}{Optimal\\ design} 
            & \Block{2-1}{Optimal\\ excitation} 
            & \Block{2-1}{\ref{sec:us_ssr}} 
            & classical & random (\citep{schiffner2019random}, \citep{schiffner_america}), optimized (\citep{cakiroglu}) \\
            & & & \gls{dl} & \citep{camsap23}\\
            & \Block{2-1}{Subsampling} 
            & \Block{2-1}{\ref{model_based_DL}} 
            & classical & \citep{perez2020subsampling} \\
            & & & \gls{dps} & \citep{huijben2020learning} \\
         \Block{18-1}{Pattern \\ recognition}
            & \Block{2-1}{Feature \\ extraction} 
            & \Block{2-1}{\ref{sec:eddy_current}, \ref{sec:3max8},\\ \ref{sec:pct}, \ref{opt._insp.}} 
            & unsupervised & \gls{pca} (\citep{RAJIC2002521}) \\
            & & & supervised & \gls{lda} (\citep{Sarg19dg}, \citep{Cy21}), \gls{svm} (\citep{Calderbank2009CompressedL}), \gls{svr} (\citep{BFL08})\\
           & \Block{6-1}{Classification} 
           & \Block{6-1}{\ref{sec:eddy_current}, \ref{sec:3max8},\\
           \ref{sec:ut_classification}, \ref{sec:optic_dl}} 
           & classical & \citep{Cy21}, \citep{davenport2007smashed} \\ 
           & & & \gls{knn} & \citep{Sarg19dg}, \citep{Cy21}\\
           & & & \gls{mlp} & \citep{KGP19} \\
           & & & Probabilistic \gls{nn} & \citep{SS99} \\
           & & & \gls{cnn} & \citep{lohit2016direct}, \citep{app13010384}, \citep{9044747}, \citep{arbaoui2021}\\
           & & & YOLOv8-cls & \citep{jocher2023yolo}\\
           & \Block{6-1}{Detection} 
           & \Block{6-1}{\ref{sec:eddy_current}, \ref{sec:3max8}, \\
           \ref{sec:optic_dl}} 
           & \gls{mlp} & \citep{RRJ02}, \citep{Sarg21}\\
           & & & \gls{rbf} & \citep{WW02}, \citep{CLH11}\\
           & & & YOLOv4 & \citep{yolov4} \\
           & & & YOLOv8 & \citep{jocher2023yolo} \\
           & & & Faster \gls{rcnn} & \citep{ren2016faster}\\
           & & & \gls{ssd} & \citep{liu2016ssd}\\
           & \Block{4-1}{Segmentation} 
           & \Block{4-1}{\ref{sec:eddy_current}, \ref{sec:optic_dl}} 
           & classical & \citep{SAHOO1988233} \\
           & & & \gls{cnn} & \citep{Defect_shape_detection}, \citep{inproceedings} \citep{semantic_seg_pedrayes}, \citep{7478072}\\
           & & & YOLOv8-seg & \citep{jocher2023yolo}\\
           & & & Mask \gls{rcnn} & \citep{he2018mask}\\
\end{NiceTabular}

%% file: sections/summary_conclusions.tex
\section{Summary and Conclusions} \label{section_summary}

In this paper we discuss the state of the art in a variety of \gls{ndt} techniques. In light of the challenges brought about by the recent advances in the corresponding industries, significant technological advances in \gls{ndt} are required. The discussion of the different \gls{ndt} modalities has emphasized some of the intricacies we are facing. More often than not, a measurement device that is sensitive for a desired physical effect will also react to unwanted effects such as environmental operating conditions (e.g., temperature) or coupling effects (e.g., sensor lift-off). The complexity of the signals requires significant experience for their interpretation in order to detect subtle effects and recognize artefacts. Training is often hindered by lack of available labeled data, which is costly to generate. Finally, since each modality is very different, generic solutions are often not available or lack in efficiency due to insufficient incorporation of the physics of the problem and the sensors.

The paper has outlined a number of AI approaches for different \gls{ndt} modalities, demonstrating examples where AI is already helpful today. In particular, we have shown that AI can assist us in detection (e.g., cracks, voids) and classification (e.g., material, defect type) as well as  segmentation problems. Moreover, AI methods have been successfully applied in the  estimation of parameters (e.g., crack depth/size, material parameters). In addition, AI methods can be a powerful tool for solving complex inverse problems for beamforming and imaging, even in complex situations such as Ultrasound tomography where full waveform inversion methods are required. Of course, they can also be used for image quality improvement (e.g., deblurring, denoising, super resolution).

Often, such methods can be tailored to be robust against calibration issues such as variations in environmental operating conditions (e.g., temperature) or sensor coupling (e.g., lift-off). The lack of training data can sometimes be alleviated through synthetic data generation for which dedicated AI methods have been devised.
An interesting thread of research is to use AI methods even very early, e.g., to assist in devising the optimal sensor placement and related design of experiment questions.

Overall, it is evident that AI methods have become indispensable tools for advancing the \gls{ndt} field further. We are certain that we are just witnessing the beginning of a research field that will give rise to even more powerful semi-automated \gls{ndt} systems which are required to keep up with the digital transformation of our industries.

%% file: sections/closing.tex
\subsection*{CRediT authorship contribution statement}

\textbf{Eduardo Pérez:} Conceptualization, Writing - Original Draft, Writing - Review \& Editing, Supervision. 
\textbf{Cemil Emre Ardic:} Writing - Original Draft. 
\textbf{Ozan Çakıroğlu:} Writing - Original Draft, Visualization. 
\textbf{Kevin Jacob:} Writing - Original Draft. 
\textbf{Sayako Kodera:} Conceptualization, Writing - Original Draft. 
\textbf{Luca Pompa:} Writing - Original Draft. 
\textbf{Mohamad Rachid:} Writing - Original Draft, Visualization, Writing - Review \& Editing. 
\textbf{Han Wang:} Writing - Original Draft, Visualization. 
\textbf{Yiming Zhou:} Writing - Original Draft, Visualization. 
\textbf{Cyril Zimmer:} Writing - Original Draft, Visualization. 
\textbf{Florian Römer:} Writing - Original Draft, Writing - Review \& Editing, Project administration, Funding acquisition. 
\textbf{Ahmad Osman:} Project administration, Funding acquisition. 

\subsection*{Conflicts of Interest}
The authors declare no conflict of interest.

\subsection*{Acknowledgements}
This work was partially supported by the Fraunhofer Internal Programs
under the grant Attract 025-601128, the Federal Ministry of Education and Research (BMBF) through the VISiMOS project with grant number 03VP10900, and the Thuringian Ministry of Economic Affairs, Science and Digital Society (TMWWDG).